\documentclass[twocolumn]{aastex62}

\newcommand{\MSTAR}{\mbox{$M_{\star}$}}
\newcommand{\RSTAR}{\mbox{$R_{\star}$}}
\newcommand{\LSOL}{\mbox{$L_{\odot}$}}
\newcommand{\RSOL}{\mbox{$R_{\odot}$}}
\newcommand{\MSOL}{\mbox{$M_{\odot}$}}
\newcommand{\TEFF}{\mbox{$T_{\rm eff}$}}

\newcommand{\VMICRO}{\mbox{$v_{\rm micro}$}}

\newcommand{\KMS}{\mbox{km s$^{-1}$}}

\newcommand{\HOH}{\mbox{H$_2$O}}

\newcommand{\corundum}{\mbox{Al$_2$O$_3$}}
\newcommand{\forsterite}{\mbox{Mg$_2$SiO$_4$}}
\newcommand{\enstatite}{\mbox{MgSiO$_3$}}

\received{}
\revised{}
\accepted{}
\submitjournal{ApJ}

\shorttitle{Three-dimensional diagnosis of the atmospheric dynamics 
of the AGB star R~Dor
}
\shortauthors{Ohnaka et al.}
\begin{document}

\title{Infrared interferometric three-dimensional diagnosis of the atmospheric 
dynamics of the AGB star R~Dor with VLTI/AMBER\footnote{Based on AMBER
  observations made with the Very Large Telescope and Very Large Telescope
  Interferometer of the European Southern Observatory. Program ID:
  092.D-0456(A), 092.D-0465(A/B)}}

\correspondingauthor{Keiichi Ohnaka}
\email{k1.ohnaka@gmail.com}

%\author[0000-0002-0786-7307]{Greg J. Schwarz}
\author{Keiichi Ohnaka}
\affiliation{Instituto de Astronom\'ia, Universidad Cat\'olica del Norte\\
Avenida Angamos 0610, Antofagasta, Chile}

\author{Gerd Weigelt}
\affiliation{Max-Planck-Insitut f\"ur Radioastronomie\\
Auf dem H\"ugel 69, 53121 Bonn, Germany}
%\collaboration{(AAS Journals Data Scientists collaboration)}

\author{Karl-Heinz Hofmann}
\affiliation{Max-Planck-Insitut f\"ur Radioastronomie\\
Auf dem H\"ugel 69, 53121 Bonn, Germany}
%\collaboration{(AAS Journals Data Scientists collaboration)}



\begin{abstract} % 250 words

The mechanism of mass loss in late evolutionary stages of low- and 
intermediate-mass stars is not yet well understood. 
Therefore, it is crucial to study the dynamics of the region within 
a few~\RSTAR, where the wind acceleration is considered to take place. 
We present three-dimensional diagnosis of the atmospheric dynamics 
of the closest asymptotic giant branch (AGB) star 
R~Dor from the low photospheric layers to the extended outer 
atmosphere---for the first time for a star other than the Sun. 
The images reconstructed with a spatial resolution of 6.8~mas---seven times 
finer than the star's angular diameter of 51.2~mas in the continuum---using 
the AMBER instrument at the 
Very Large Telescope Interferometer show a large, bright region over the 
surface of the star and an extended atmosphere. 
The velocity-field maps over the star's surface and atmosphere
obtained from the Mg and \HOH\ lines near 
2.3~\micron\ forming at atmospheric heights below $\sim$1.5~\RSTAR\ 
show little systematic motion beyond the measurement uncertainty of 1.7~\KMS. 
In marked contrast, the velocity-field map obtained from the CO first 
overtone lines reveals systematic outward motion at 7--15~\KMS\ in 
the extended outer atmosphere at a height of $\sim$1.8~\RSTAR. 
Given the detection of dust formation at $\sim$1.5~\RSTAR, 
the strong acceleration of material between $\sim$1.5 and 1.8~\RSTAR\ 
may be caused by the radiation pressure on dust grains. 
However, we cannot yet exclude the possibility that 
the outward motion may be intermittent, 
caused by ballistic motion due to convection and/or pulsation. 

\end{abstract}



\section{Introduction}
\label{sect_intro}

Mass loss in cool evolved stars---stars in the red giant branch (RGB) and 
asymptotic giant branch (AGB) as well as red supergiants---affects 
not only the stellar evolution but also the chemical enrichment of galaxies. 
Despite such importance, the mass-loss mechanism in cool evolved stars 
is a long-standing problem over half a century since the discovery of 
the mass outflow in the red supergiant $\alpha$~Her by \citet{deutsch56}. 
In case of AGB stars, 
it is often argued that 
the levitation of the atmosphere by pulsation 
leads to dust formation, and the radiation pressure on dust grains may drive 
slow but dense stellar winds \citep[e.g.,][]{hoefner18}. 
In case of oxygen-rich AGB stars, the latest theoretical models 
suggest that scattering---instead 
of absorption---of stellar photons by large (a few 0.1~\micron), 
transparent grains may be the driving force \citep{hoefner08,hoefner16}. 

To improve our understanding of the mass-loss mechanism, it is crucial 
to study the region within a few~\RSTAR, where the wind acceleration is 
considered to take place. Given that even the largest angular diameter 
of nearby AGB stars is mere 50~mas (that of R~Dor, the target of the 
present paper), 
we need milliarcsecond spatial resolution to resolve the key region. 
Optical and infrared interferometric techniques enable us to 
spatially resolve not only the extended atmosphere but also inhomogeneous 
surface structures in AGB stars
\citep[e.g.,][]{ragland08,lacour09,lebouquin09,wittkowski17,paladini18}. 
Some of the high-spatial resolution images 
reveal the presence of inhomogeneous 
molecular outer atmospheres extending out to a few~\RSTAR. 

The dust formation region close to the star has also been spatially resolved. 
Optical long-baseline interferometric polarimetric observations of 
oxygen-rich AGB stars by 
\citet{ireland05} suggest the scattering of stellar light by dust forming 
closer than about 3~\RSTAR. 
The near-infrared aperture-masking experiments of \citet{norris12} 
show the presence of $\sim$0.3~\micron -sized transparent grains at 
$\sim$1.5~\RSTAR\ in three AGB stars, including the target of the present 
paper R~Dor. 
More recently, visible polarimetric imaging observations have revealed 
the formation of clumpy dust clouds around several oxygen-rich AGB stars 
\citep{khouri16,khouri18,ohnaka16,ohnaka17a,adam19}. 
In particular, the polarimetric images and the modeling of the nearby, 
well-studied AGB stars R~Dor and W~Hya \citep{khouri16,ohnaka16,ohnaka17a} 
show that dust forms as close as 
at $\sim$1.5~\RSTAR, consistent with the results of \citet{norris12}. 

It is of paramount importance to study the dynamics of the 
extended atmosphere for clarifying the mechanism responsible 
for the stellar wind acceleration. Detailed spectroscopic analyses of 
near-infrared atomic and molecular lines show distinct gas motion 
in the photospheric layers and the extended outer atmosphere 
\citep[e.g.,][]{hinkle79}. Nevertheless, it is not straightforward to 
ascertain the atmospheric (i.e., geometrical) height of different layers 
from the spatially unresolved spectroscopic data. Furthermore, the 
inhomogeneities over the surface and atmosphere seen in the high-spatial 
resolution images complicate the interpretation. 
A direct approach is to measure the line-of-sight velocity at each position 
over the spatially resolved surface and atmosphere of stars. 
High-spatial and high-spectral resolution infrared interferometric 
observations of the 2.3~\micron\ CO lines enabled us to 
obtain a two-dimensional map of the line-of-sight 
velocity over the surface and atmosphere of the red supergiant 
Antares---for the first time for a star other than the Sun. 
In the present paper, we extend this approach to three-dimensional 
diagnosis of the atmospheric dynamics for the AGB star R~Dor. 

Our observations and data reduction are summarized in Section~\ref{sect_obs}. 
The analysis of the reduced data 
is described in Section~\ref{sect_analysis}. We present the results in 
Section~\ref{sect_results} followed by discussion and conclusion in 
Section~\ref{sect_discuss}. 

%% The "ht!" tells LaTeX to put the figure "here" first, at the "top" next
%% and to override the normal way of calculating a float position
\begin{figure}
\epsscale{1.18}
\plotone{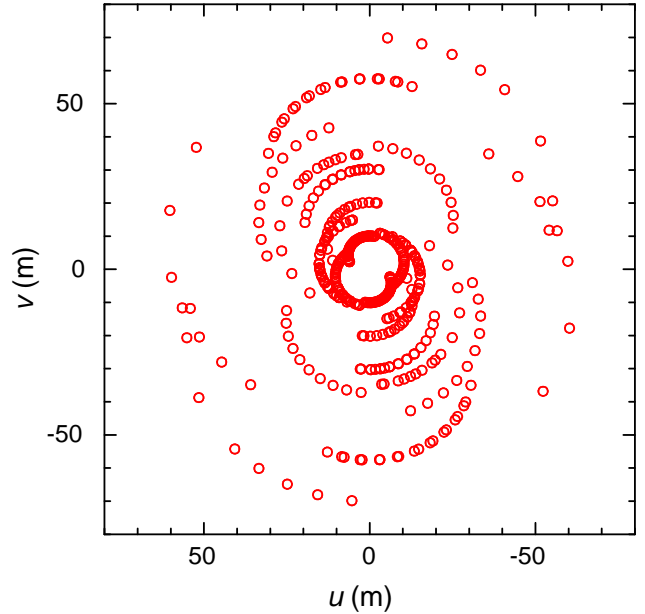}
\caption{
The $uv$ coverage of our VLTI/AMBER observations of R~Dor with six 
different AT configurations. 
\label{rdor_uv}
}
\end{figure}

\section{Observations and data reduction}
\label{sect_obs}

\subsection{AMBER observations}
\label{subsect_obs}

Our target R~Dor is one of the well-studied AGB stars, 
classified as a semiregular variable of SRB type with a spectral type of M8III 
\citep{samus17}. 
At a distance of 55~pc \citep[Hipparcos parallax =
  18.31~mas,][]{vanleeuwen07}\footnote{GAIA parallax is not available for R~Dor.}, 
it is the closest AGB star. 
Its extremely high brightness in the infrared 
and its large angular diameter of 51.2~mas 
(see Sect.~\ref{subsect_lddfit}) provide 
an excellent opportunity to not only obtain detailed images of its surface 
and atmosphere but also velocity-field maps (i.e., maps of the line-of-sight 
velocity) of the atmospheric gas motion. 

We observed R~Dor with the near-infrared spectro-interferometric instrument 
AMBER \citep{petrov07} at the Very Large Telescope Interferometer (VLTI) of the 
European Southern Observatory. 
Our AMBER observations (Program ID: 092.D-0456(A), 092.D-0465(A/B), 
P.I.: K.~Ohnaka) took place on seven nights allocated 
within 10 days, 2013 December 1, 2, 6, 7, 8, 9, and 10 UT, thus 
minimizing possible variations in the structures over the surface and 
atmosphere. We used six array configurations of the Auxiliary Telescopes 
(ATs), A1-B2-D0, A1-B2-C1, A1-C1-D0, B2-C1-D0, D0-G1-H0, and G1-H0-I1. 
The intensive observing program provided a wide baseline coverage from 
6.5 to 70~m.  
This baseline coverage and the southern declination of R~Dor ($-62^\circ$) 
allowed us to obtain a good $uv$ coverage as shown in 
Figure~\ref{rdor_uv}, with a spatial resolution of 6.8~mas. 

The AMBER instrument combines this milliarcsecond angular resolution with 
a high spectral resolution of up to 12\,000, which is needed to extract 
dynamical information over the surface and atmosphere of stars 
using Doppler shifts of individual spectral lines. 
We observed the wavelength range between 2.278 and 2.308~\micron\ with 
the spectral resolution of 12\,000. 
In this spectral window, there are strong CO first overtone lines of the 
$v = 2 - 0$ transition that form in 
the {\em extended outer atmosphere} as well as 
other molecular and atomic lines, such as due to \HOH, HF, Ti, and Mg, 
which form in {\em deeper atmospheric layers}. 
Therefore, these distinct spectral lines allow us to probe the gas dynamics 
at different atmospheric heights.
The VLTI fringe tracker FINITO was not used because it was saturated by 
the high brightness of R~Dor. 

We observed Canopus ($\alpha$~Car, spectral type A9II) for the calibration 
of the interferometric data and also for the spectroscopic calibration. 
We adopted the angular diameter of $6.93 \pm 0.15$~mas 
based on fitting with a linearly
limb-darkened disk \citep{domiciano08}. A journal of our observations is 
given in Table~\ref{obslog}.

\subsection{Data reduction}
\label{subsect_reduction}

The recorded spectrally dispersed interferograms were processed with 
amdlib ver.~3.0.7\footnote{http://www.jmmc.fr/data\_processing\_amber.htm}, 
which outputs the squared visibility amplitude, closure phase, 
and wavelength-differential phase, together with the spatially unresolved 
flux (i.e., spectrum) as a function of wavelength based on the P2VM 
algorithm \citep{tatulli07,chelli09}. Before processing with amdlib, 
all the raw data were spectrally binned to a spectral resolution of 8\,000 
with a running box filter to increase S/N sufficiently high for the image 
reconstruction as described in \citet{ohnaka09,ohnaka11,ohnaka13}. 
The interferometric observables (squared visibility amplitude, 
closure phase, and wavelength-differential phase) of each data set were 
computed by taking the average of all frames ($N_{\rm f}$), discarding 
20\% of the frames with the lowest fringe S/N. 

We used the telluric lines observed in the spectrum of Canopus for the 
wavelength calibration. The observed spectrum of Canopus was first 
spline-interpolated, and the pixel positions ($i$) of the identified telluric 
lines were fitted with a quadratic function, 
$\lambda = a + b \, i + c \, i^2$. 
The uncertainty of the wavelength calibration is $7.7\times 10^{-6}$~\micron\ 
(1.0~\KMS). 

The spectrum of R~Dor can be spectroscopically calibrated with that of 
Canopus as 
$F_{\rm sci}^{\rm true} = F_{\rm sci}^{\rm obs} \times F_{\rm cal}^{\rm
  true}/F_{\rm cal}^{\rm obs}$, 
where $F_{\rm sci (cal)}^{\rm true}$ and $F_{\rm sci (cal)}^{\rm obs}$
denote the true and observed spectra of the science target (R~Dor) or 
the calibrator (Canopus), respectively. 
A spectroscopically calibrated high-resolution spectrum of Canopus can be 
used as $F_{\rm cal}^{\rm  true}$.  
However, we found none in the literature for the observed wavelength region. 
Therefore, we used as a proxy of Canopus the high-resolution spectrum of 
HD6130 obtained by \citet{park18}, 
because the spectral type of HD6130, F0II, is close to that of Canopus. 
The spectrum of HD6130 originally obtained 
with a spectral resolution of 45\,000 was convolved to the spectral resolution 
of 8\,000 of our AMBER data before the spectroscopic calibration 
of R~Dor.

\begin{figure*}
\epsscale{1.16}
\plotone{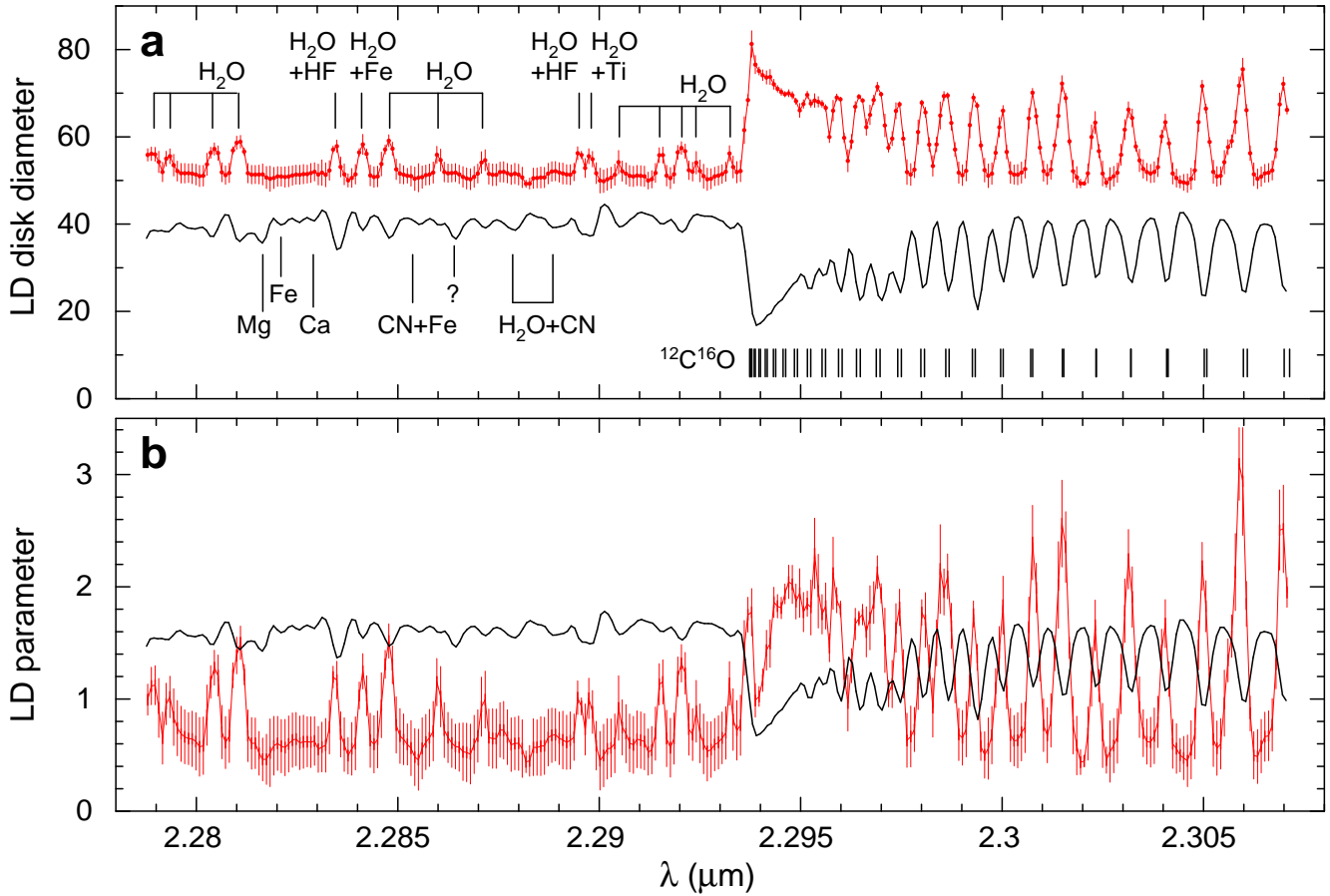}
\caption{
Power-law-type limb-darkened (LD) disk fitting of the AMBER data of R~Dor.
{\bf a:} Limb-darkened disk diameter (red line) and the scaled observed 
spectrum (black line).
The CO and \HOH\ lines that give rise to the increases in the limb-darkened 
disk diameter and the limb-darkening parameter are marked. 
Other atomic and molecular lines without a increase 
in the limb-darkened disk diameter are also marked. 
While the wavelength of the line at 2.2864~\micron\ coincides with a cobalt 
line, it is possible that it is blended with an unidentified line as discussed 
in \citet{ohnaka19}. 
{\bf b:} Limb-darkening parameter (red line) and the scaled observed spectrum 
(black line). 
\label{rdor_lddfit}
}
\end{figure*}

\section{Analysis}
\label{sect_analysis}

\subsection{Limb-darkened disk fitting}
\label{subsect_lddfit}

Fitting the observed visibilities with a limb-darkened disk is useful for 
obtaining an approximate size of the object as a function of wavelength. 
The best-fit limb-darkened disk model can also be used as the initial model 
for the image reconstruction. 
Figure~\ref{rdor_lddfit} shows the results 
of fitting with a power-law-type limb-darkened disk \citep{hestroffer97}. 
We estimated the errors in two parameters---limb-darkened disk diameter and 
limb-darkening parameter---at each wavelength channel using the bootstrap 
method \citep{efron93}. 
Then we took the mean of these parameters at the continuum spectral 
channels (avoiding the \HOH\ lines that give rise to the increases in the 
limb-darkened disk diameter) as representative values in the continuum. 
Because the errors are dominated by the systematic 
errors in the absolute calibration of the visibilities, the errors were 
not reduced by a factor of $1/\sqrt{N_{\rm cont}}$, where $N_{\rm cont}$ is 
the number of the continuum spectral channels. Instead we took the mean of 
the errors at the continuum channels. 
We obtained a limb-darkened disk diameter and a limb-darkening parameter 
of $51.18 \pm 2.24$~mas and $0.61 \pm 0.24$, respectively, in the continuum. 
The angular size is in fair agreement with the uniform-disk diameter of 
$57 \pm 5$~mas measured at 1.25~\micron\ by \citet{bedding97}. 

In the CO band head at 2.294~\micron\ and individual CO lines, the 
limb-darkened disk diameter increases up to $\sim$80~mas, while the 
limb-darkening parameter increases up to 2--3. 
However, 
the reduced $\chi^2$ of the fitting is 10--30 in the continuum and 
as large as 70 in the CO lines. This indicates the presence of surface 
structures in the continuum and more complex structures in the lines, 
which indeed manifest themselves as the large, bright region over the 
surface and the extended atmosphere in the reconstructed images as 
presented in Section~\ref{sect_results}.

It is worth noting that some spectral lines shortward of the CO band head 
give rise to spike-like increases in the limb-darkening diameter as well as in 
the limb-darkening parameter, even though they appear to be weak in the 
spatially unresolved (i.e., spatially averaged over the stellar 
disk and the atmosphere) spectrum. 
As identified in Figure~\ref{rdor_lddfit}, most of 
these lines are due to \HOH, some of which are blended with other atomic 
or molecular lines. This was reported in the visibility analysis of the 
M7 giant SW~Vir \citep{ohnaka19} and modeled by the \HOH\ layer extending 
out to $\sim$2~\RSTAR. The \HOH\ lines appear to be weak in the 
spatially unresolved spectrum, because the 
\HOH\ absorption expected over the stellar disk is filled in by the emission 
due to the same \HOH\ lines originating from the extended \HOH\ layer. 
As we present in Section~\ref{subsect_res_images}, the extended \HOH\ layer 
manifests itself in the images reconstructed in these \HOH\ lines.

\begin{figure}
\epsscale{1.2}
\plotone{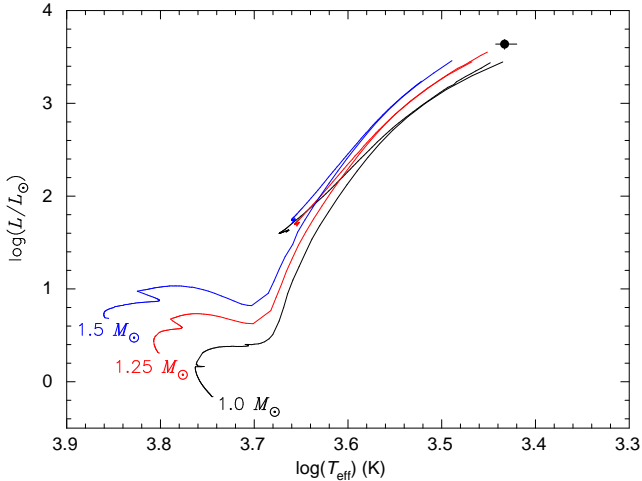}
\caption{
H-R diagram with the position of R~Dor (filled circle) and evolutionary 
tracks with $Z = 0.014$ for 1, 1.25, and 1.5~\MSOL\ stars from 
\citet{lagarde12}. 
\label{hr_diagram}
}
\end{figure}

\begin{deluxetable}{lrr}
\tablecaption{Basic parameters of R~Dor \label{tab_rdor_param}}
\tablecolumns{3}
%\tablenum{2}
\tablewidth{0pt}
\tablehead{
\colhead{} &
\colhead{Measured} &
\colhead{MARCS model}
}
\startdata
Distance (pc) & $55\pm3^{a}$ & --- \\
Bolometric flux (W~m$^{-2}$) & $ 4.71 \times 10^{-8} $ & --- \\
Luminosity (\LSOL) & 4350 $\pm$ 520 & 1326 \\
\TEFF\ (K) & 2710 $\pm$ 70 & 2700 \\
Radius (\RSOL) & 298 $\pm$ 21 & 165 \\
$\log g$ (cm~s$^{-2}$) & $-0.6 \pm 0.1$ & 0.0 \\ 
Initial mass (\MSOL) & 1--1.25 & --- \\
Current mass \MSTAR\ (\MSOL) & 0.7--1.0 & 1 \\
\VMICRO\ (\KMS) & 2.0$^{b}$ & 2.0 \\
\mbox{[Fe/H]} & --- & 0.0 \\
Chemical composition & --- & moderate-CN \\
Age (Gyr) & 6--14 & --- \\
\enddata
\tablenotetext{a}{Based on the parallax from \citet{vanleeuwen07}.}
\tablenotetext{b}{Adoped from \citet{ryde02}.}
\end{deluxetable}

\subsection{Hydrostatic photospheric model}
\label{subsect_marcs}

As a reference for the measurement the line-of-sight velocity from spectral 
lines in spatially resolved spectra presented 
in Section~\ref{subsect_res_velmap}, 
we used a synthetic spectrum computed 
from the spherical MARCS model atmosphere \citep{gustafsson08} 
with the parameters appropriate for R~Dor. 
The MARCS model atmospheres are specified by 
effective temperature, surface gravity, micro-turbulent velocity, 
chemical composition, and stellar mass. We determined these 
parameters as follows. 

We first calculated the bolometric flux by collecting the photometric 
and spectrophotometric data from the visible to the 
mid-infrared \citep{mermilliod87,ducati02,stewart15,sloan03}. 
A visible extinction $A_{V}$ of 0.090 
was derived from the three-dimensional 
map of the interstellar extinction presented by \citet{arenou92}. 
The collected (spectro)photometric data were then de-reddened by applying 
the wavelength-dependence of the interstellar extinction of 
\citet{cardelli89}. 
The bolometric flux was derived to be $4.71 \times 10^{-8}$~W~m$^{-2}$. 
We estimated the uncertainty in the bolometric flux to be 5\% based on the 
infrared light curves at 1.25, 2.2, 3.5, and 4.9~\micron\ reported by 
\citet{price10}. 
Since the majority of the flux is emitted between 1 and 5~\micron, 
the variability at shorter or longer wavelengths does not noticeably affect 
the estimate of the bolometric flux. 
Combined with the measured limb-darkened disk diameter of 
51.18$\pm$2.24~mas, 
we obtained an effective temperature (\TEFF) of $2710\pm 70$~K. 

The derived luminosity is $4350\pm 520$~\LSOL\ 
based on the bolometric 
flux and the measured parallax of $18.31 \pm 0.99$~mas \citep{vanleeuwen07}. 
Figure~\ref{hr_diagram} shows a 
comparison of the position of R~Dor on the HR-diagram with 
the theoretical evolutionary tracks of 1, 1.25, and 1.5~\MSOL\ stars with 
the solar metallicity ($Z = 0.014$) from \citet{lagarde12}. 
While none of these theoretical evolutionary tracks reaches the luminosity 
of R~Dor, those with the initial stellar masses of 1--1.25~\MSOL\ 
are the closest to the observed value. 
This suggests an age of 14~Gyr and 6~Gyr with a current stellar mass of 
0.7~\MSOL\ and 1.0~\MSOL\ for the initial mass of 1 and 1.25~\MSOL, 
respectively. Therefore, the star has lost $\sim$0.3~\MSOL\ in either case. 
The stellar radius estimated from the measured limb-darkened disk diameter 
xsand distance is $ 298\pm 21$~\RSOL. 
Then the surface gravity is estimated to be 
$\log g  = -0.6 \pm 0.1$. 
We adopted a micro-turbulent velocity (\VMICRO) 
of 2~\KMS\ as in the previous analysis 
of the infrared spectrum of R~Dor by \citet{ryde02}. 

The MARCS model with the parameters closest to those of R~Dor is 
specified by \TEFF\ = 2700~K, $\log g = 0.0$, \MSTAR\ = 1~\MSOL, 
and \VMICRO\ = 2~\KMS. 
We assumed [Fe/H] = 0.0 and 
the ``moderately CN-cycled'' chemical composition of the 
MARCS model grid, which is appropriate for oxygen-rich AGB stars. 
The derived basic parameters of R~Dor, together with those of the 
best MARCS model for R~Dor, are summarized in Table~\ref{tab_rdor_param}. 
The luminosity and the stellar radius 
of the MARCS model are significantly lower and smaller than those 
observationally derived. This is because the surface gravity of the best MARCS 
model is still smaller than the observationally estimated value (no MARCS 
model with $\log g \approx -0.6$ is available for the \TEFF\ of R~Dor). 
However, the model spectrum is primarily used to 
ascertain the wavelengths of the spectral lines for the measurement of 
the line-of-sight velocity, not the line strengths, 
which may be affected by the difference in the surface gravity. 
Therefore, the measurement of the line-of-sight velocity over the reconstructed 
images presented in Section~\ref{subsect_res_velmap} 
is little affected by the difference in the surface gravity. 

In the calculation of the synthetic spectrum in the observed wavelength 
region, we included the lines of 
$^{12}$C$^{16}$O, $^{12}$C$^{17}$O, $^{12}$C$^{18}$O, 
$^{13}$C$^{16}$O, $^{1}$H$_2^{16}$O, $^{12}$C$^{14}$N, $^{13}$C$^{14}$N, and 
$^{1}$H$^{19}$F, as well as atomic lines based on the line lists available in 
the literature
\citep{goorvitch94,polyansky18,sneden14,jorissen92,kurucz95}\footnote{\HOH :
  http://www.exomol.com/data/molecules/H2O\\/1H2-16O/POKAZATEL/\\
CN: https://www.as.utexas.edu/\~{}chris/lab.html\\
Atomic lines: https://www.cfa.harvard.edu/amp/ampdata\\
/kurucz23/sekur.html}. 
The synthetic spectrum was convolved to the spectral resolution of 8\,000 
of the spectrally binned AMBER data. Then it was shifted in wavelength from 
the laboratory frame to the observed frame, accounting for 
the Earth's motion at the time of the observations and the systemic velocity 
of R~Dor. 
Since spectral lines in the visible and near-infrared are affected by stellar 
pulsation and convection \citep[e.g.,][]{hinkle02}, they are not optimal for 
deriving the systemic velocity of the star. Instead, radio and far-IR 
observations of molecular lines originating from approximately 
spherically expanding circumstellar envelopes 
are often used to obtain the systemic velocity. 
The values for R~Dor in the literature range from 6 to 8~\KMS\ 
in the local standard 
of rest \citep{gonzalez_delgado03,decin18,homan18}. 
In the present work, we adopted the 7.5~\KMS\ recently derived by 
\citet{vandesande18} and \citet{homan18}, 
which translates into a heliocentric velocity of 23.9~\KMS. 

\begin{figure*}
\epsscale{1.18}
\plotone{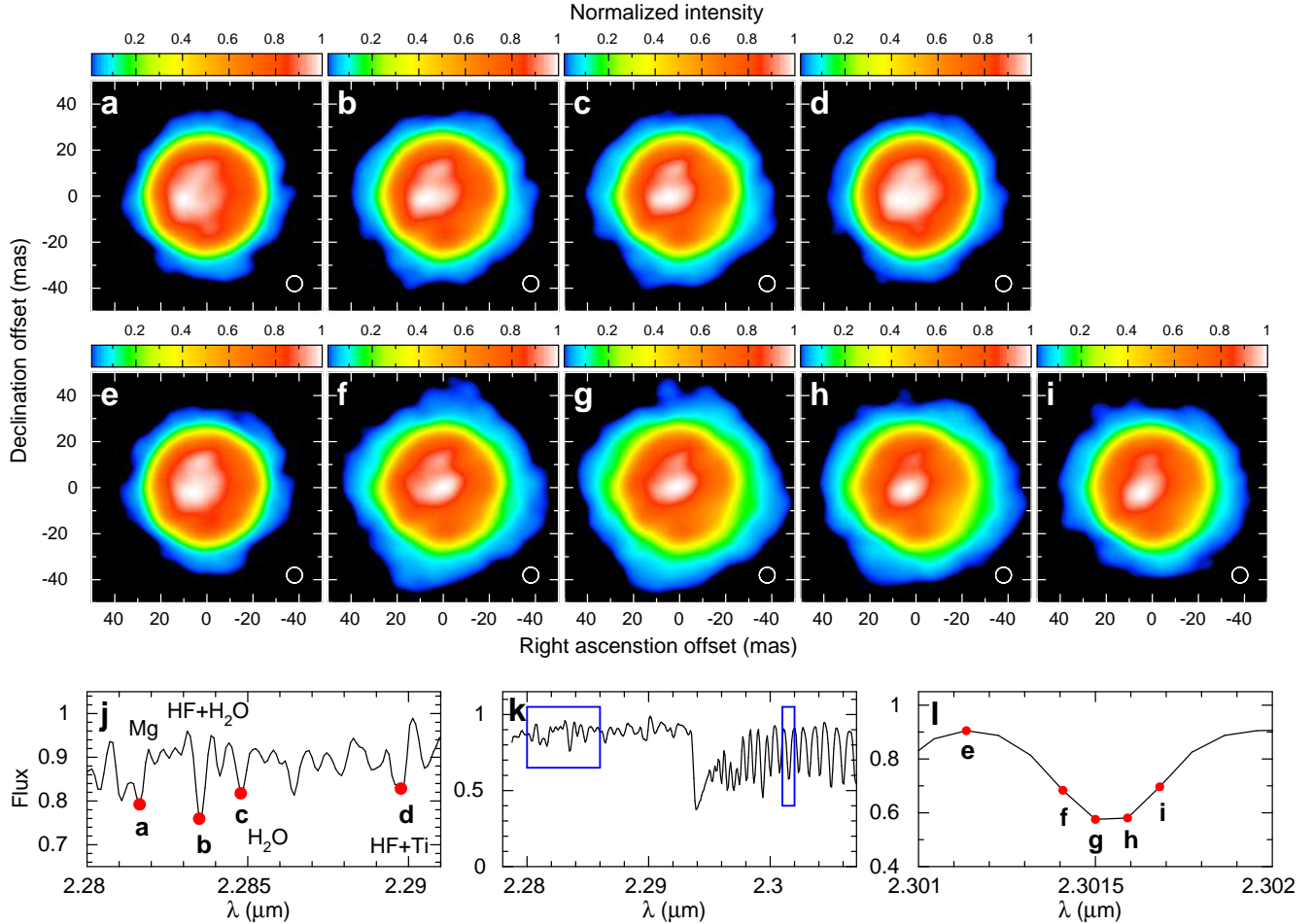}
\caption{
Wavelength-dependent images of the surface and extended atmosphere of R~Dor. 
The images reconstructed at nine representative 
wavelength channels in the molecular and atomic lines of Mg, Ti, HF, and 
\HOH\ (panels {\bf a--d}) 
and across one of the CO lines (panels {\bf e--i}) are shown. 
Each image is normalized with the maximum intensity. 
The color scale is set so that the intensity lower than 1\% of the 
maximum intensity is cut off (appearing in black), 
to show the images used for the derivation of the line-of-sight velocity 
maps as described Section~\ref{subsect_res_velmap}.
The image in the continuum ({\bf e}) shows a large, bright region over the 
stellar disk, which can also be seen in the images in the molecular and 
atomic lines. It should be noted, however, that the intensity contrast 
differs in the continuum ($\sim$25\%) and in the lines ($\sim$40\%). 
The extended atmosphere outside the stellar disk 
can be seen in the images obtained in the \HOH\ 
lines ({\bf b} and {\bf c}), and appears more pronounced in the images 
obtained in the CO line ({\bf f}, {\bf g}, and {\bf h}). 
The beam size 
($6.8 \times 6.8$~mas) is shown in the lower right corner of each panel. 
North is up, east to the left. 
Panels {\bf j--l:} Spectrum of R~Dor 
in the entire spectral range of our observations is shown 
in {\bf k}. Enlarged views of two wavelength ranges 
(marked with the rectangles in panel {\bf k}) 
are shown in panels {\bf j} and {\bf l}, where 
the wavelength channels of the images in panels {\bf a--i} are marked with 
the corresponding alphabetic characters. 
\label{rdor_images}
}
\end{figure*}

\subsection{Image reconstruction}
\label{subsect_reconst}

We reconstructed the image at each wavelength channel using MiRA 
ver~0.9.9\footnote{http://cral.univ-lyon1.fr/labo/perso/eric.thiebaut/?Software/
  \\ MiRA} \citep{thiebaut08}. 
We applied the technique to restore the Fourier phases from 
wavelength-differential phase measurements 
\citep{petrov07,schmitt09,millour11,ohnaka11,ohnaka13,weigelt16,ohnaka17b}. 
The restored Fourier phases 
were used for the image reconstruction together 
with the squared visibility amplitudes and closure phases. 
We also carried out reconstruction experiments with simulated data 
to examine the reliability of the image reconstruction from the observed
data. 
Details of the image reconstruction are described in 
Appendix~\ref{appendix_reconst}. 

The reconstructed 
images were convolved with a two-dimensional Gaussian beam with a full width 
at half maximum (FWHM) of $\lambda/B_{\rm max}$ = 6.8~mas, 
where $\lambda$ was taken to be the central 
wavelength of the observed spectral window (2.3~\micron), and 
$B_{\rm max}$ is the maximum projected baseline length of our observations 
(70~m). 
As presented in Appendix~\ref{appendix_reconst}, we carried out the image 
reconstruction both with the quadratic (Tikhonov) regularization and the 
maximum entropy regularization with different reconstruction parameters. 
We computed the median from the images 
reconstructed with different reconstruction parameters at each wavelength 
channel.

\begin{figure*}
\epsscale{1.16}
\plotone{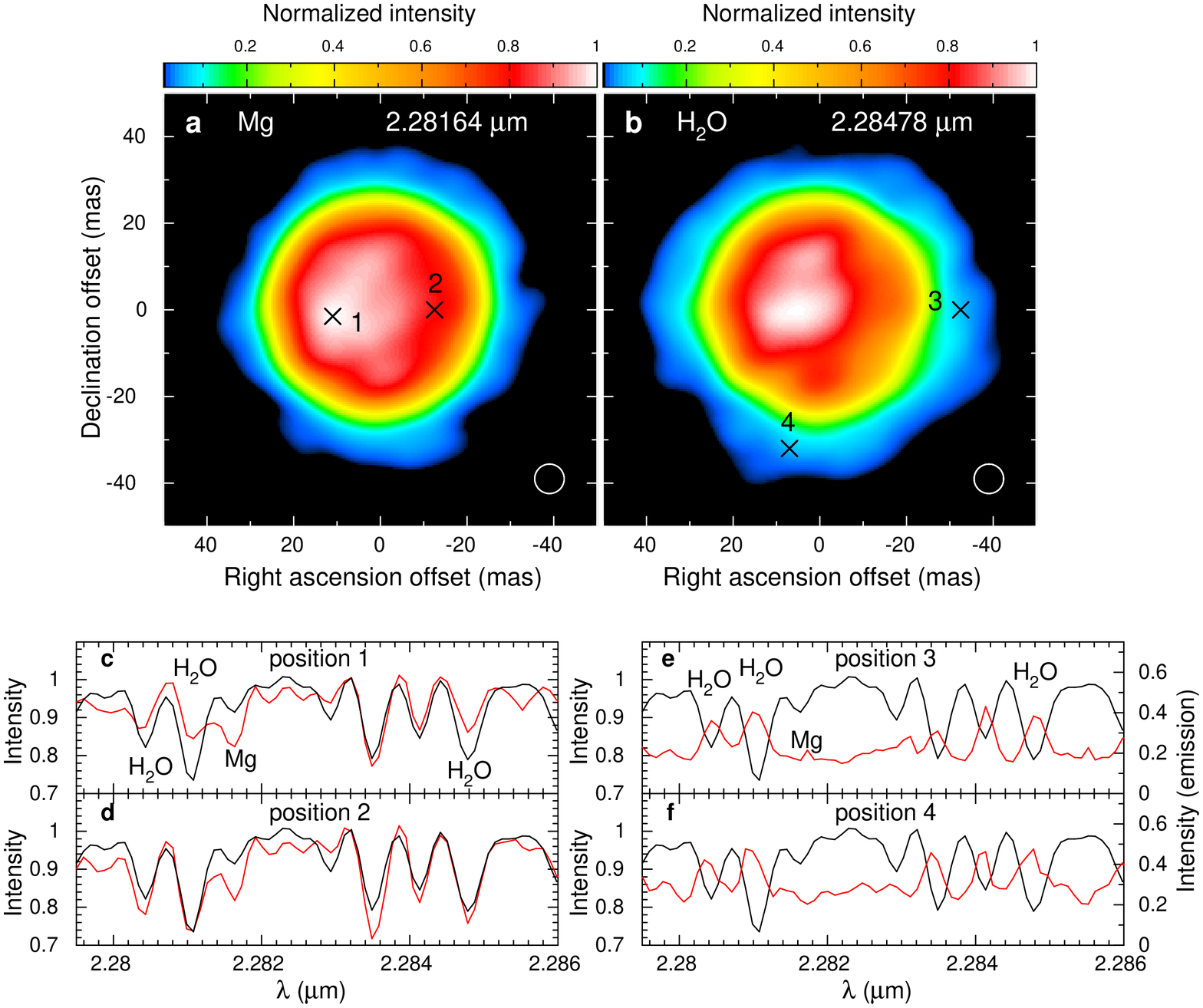}
\caption{
Spatially resolved spectra of the lines of Mg and \HOH\ 
obtained over the surface and atmosphere of R~Dor. 
The Mg and \HOH\ lines probe the gas dynamics at atmospheric heights of 
$\la$1.13~\RSTAR\ and $\sim$1.5~\RSTAR, respectively. 
{\bf a} and {\bf b:} The images reconstructed at the center of the Mg line 
at 2.28164~\micron\ and the \HOH\ line at 2.28478~\micron, respectively. 
The positions 1, 2, 3, and 4 are where the spatially resolved spectra 
exemplarily shown in {\bf c}--{\bf f} were extracted. 
North is up, east is to the left. 
{\bf c}--{\bf f:} Spatially resolved spectra obtained at the positions 1, 2, 
3, and 4 are shown by the red lines. The emission line spectra at positions 
3 and 4 are scaled, as shown in the right abscissa, to facilitate 
the visual inspection. 
The black lines show the synthetic spectrum computed from the 
spherical hydrostatic photospheric model without atmospheric motion. 
The Mg and \HOH\ lines used to derive the maps of the line-of-sight velocity 
(Figures~\ref{rdor_velmap}a and \ref{rdor_velmap}b) are identified. 
The spatially resolved spectra of these lines obtained at the positions 
1, 2, and 3 (panels {\bf c}, {\bf d}, and {\bf e}) do not show clear 
wavelength shifts with respect to the hydrostatic model spectrum, 
in contrast to the CO line spectra shown in Figure~\ref{rdor_specpos_CO}. 
Only in the spectrum at the position 4 (panel {\bf f}) 
do the \HOH\ lines show some blueshifts. 
\label{rdor_specpos_weak}
}
\end{figure*}

%\newpage
\section{Results}
\label{sect_results}

\subsection{Wavelength-dependent images of the surface and extended atmosphere}
\label{subsect_res_images}

Figure~\ref{rdor_images} shows images of R~Dor reconstructed at nine 
representative wavelength channels in different spectral lines as well as 
in the continuum. The image reconstructed in the continuum 
(Figure~\ref{rdor_images}e) shows a large, irregularly shaped bright 
region over the stellar surface, which is well resolved with the angular 
resolution of 6.8~mas---seven times finer than the star's angular diameter of 
51.2~mas. 
The intensity contrast of the bright region with respect to the surrounding 
region is about 25\%. 
This is very different from the surface of the red supergiant 
Antares seen in the continuum, which only exhibits a weak, large spot 
with an intensity contrast of 3--4\% \citep{ohnaka17b}. 
The presence of a high-contrast bright region is qualitatively 
consistent with the latest three-dimensional convection simulations of AGB 
stars \citep{freytag17}. 
As will be discussed in Section~\ref{sect_discuss}, dust formation is 
detected at $\sim$1.5~\RSTAR\ toward R~Dor \citep{norris12,khouri16}. 
However, the contribution of dust in the observed continuum images is likely 
to be negligible, 
because the flux contribution of dust-scattered light is estimated 
to be only $\sim$1.4\% at 2.06~\micron\ for R~Dor \citep{norris12}.

The image reconstructed in the Mg line (Figure~\ref{rdor_images}a) 
appears to be 
nearly the same as the continuum image without a noticeable trace of an 
extended atmosphere. 
This suggests that the Mg line originates in deep layers 
(i.e., close to the star), whose geometrical thickness is 
not spatially resolved with the resolution of 6.8~mas. 
Therefore, the line probes atmospheric heights lower than 
$\sim$1.13~\RSTAR\ (measured from the stellar center). 
However, the images reconstructed in the 
\HOH\ lines shown in Figures~\ref{rdor_images}b and \ref{rdor_images}c 
(the former line blended with an HF line) show an extended atmosphere 
with a radius of $\sim$38 mas ($\sim$1.5~\RSTAR). 
Moreover, 
the CO line images (Figures~\ref{rdor_images}f--\ref{rdor_images}i) show 
an atmosphere even more extended out to $\sim$1.8~\RSTAR. 
The intensity contrast of the bright region over the surface increases 
from the 25\% in the continuum to $\sim$40\% in the \HOH\ and CO line images. 
It may represent a region with higher temperatures 
throughout the atmosphere. 

We note that the hydrostatic MARCS photospheric model with the parameters 
appropriate for R~Dor extends only to 1.12~\RSTAR. 
As mentioned in Section~\ref{subsect_marcs}, the surface gravity of the 
best MARCS model ($\log g = 0.0$) is still smaller than the observationally 
estimated value ($\log g = -0.6$). 
Since the geometrical thickness of the atmosphere with respect 
to the stellar radius, $\Delta r/\RSTAR$, is proportional to $g^{-1/2}$ 
\citep{joergensen92}, the geometrical thickness of the MARCS photospheric 
model after correcting for the difference in the surface gravity is 0.24. 
This means that the photosphere extends only to 1.24~\RSTAR\ at most. 
Therefore, the extended outer atmosphere seen in the \HOH\ and CO line images 
cannot be explained by the hydrostatic model atmosphere. 
Such extended atmospheres in Mira-type AGB stars are fairly explained 
by large-amplitude stellar pulsation 
\citep[e.g.,][]{wittkowski08,wittkowski11,hillen12}. 
However, R~Dor is a semiregular variable, and its variability amplitude is 
$\Delta V \approx 1.5$~mag based on the light curve compiled by the 
American Association of Variable Star Observers (AAVSO). 
This is much smaller than that of typical Mira-type AGB stars. 
The presence of the extended atmosphere 
has been detected in other non-Mira type AGB stars 
by infrared-interferometric observations in the CO and 
\HOH\ lines but without images so far 
\citep[e.g.,][]{marti-vidal11,ohnaka12,ohnaka19}. 
Our \HOH\ and CO line images of R~Dor are the first images of the 
extended atmosphere of a non-Mira AGB star.

\subsection{Spatially resolved spectra over the surface and extended atmosphere}
\label{subsect_res_spatspec}

The images reconstructed at 309 wavelength channels across the CO lines 
and other spectral lines allowed us to extract the spatially 
resolved spectrum at each position over the surface and atmosphere of the 
star. 
For this purpose, 
the intensity of the image reconstructed and convolved with the beam 
at each wavelength channel was scaled so that the flux 
integrated over the entire image is equal to the flux observed in the 
spectroscopically calibrated spectrum of R~Dor.

Figure~\ref{rdor_specpos_weak} shows the spatially resolved spectra 
across the lines of Mg and \HOH\ (red lines) 
obtained at four different positions 
marked in Figures~\ref{rdor_specpos_weak}a and \ref{rdor_specpos_weak}b. 
Also shown is the synthetic spectrum (black lines) computed from the 
hydrostatic MARCS photospheric model (without atmospheric motion) 
with the parameters appropriate for R~Dor as described in
Section~\ref{subsect_marcs}. 
At the positions over the stellar surface, the lines in the spatially resolved 
spectrum appear in absorption 
(Figures~\ref{rdor_specpos_weak}c and \ref{rdor_specpos_weak}d). 
However, 
the \HOH\ lines at the positions 3 and 4 over the atmosphere turn into 
emission (Figures~\ref{rdor_specpos_weak}e and \ref{rdor_specpos_weak}f), 
as expected from Kirchhoff's law. The Mg line at 
2.28164~\micron\ does not appear in emission at the positions 3 and 4, 
because as seen in Figures~\ref{rdor_images}a and \ref{rdor_images}e 
the extension of the Mg line image is nearly the same as the continuum image. 
It is worth noting, compared to the 
CO lines described below, that the Mg and \HOH\ lines appear approximately 
at the wavelengths predicted by the hydrostatic model without significant 
wavelength shifts, except for the position 4 where the \HOH\ emission lines 
show some blueshifts (Figure~\ref{rdor_images}f).

\begin{figure*}
\epsscale{1.1}
\plotone{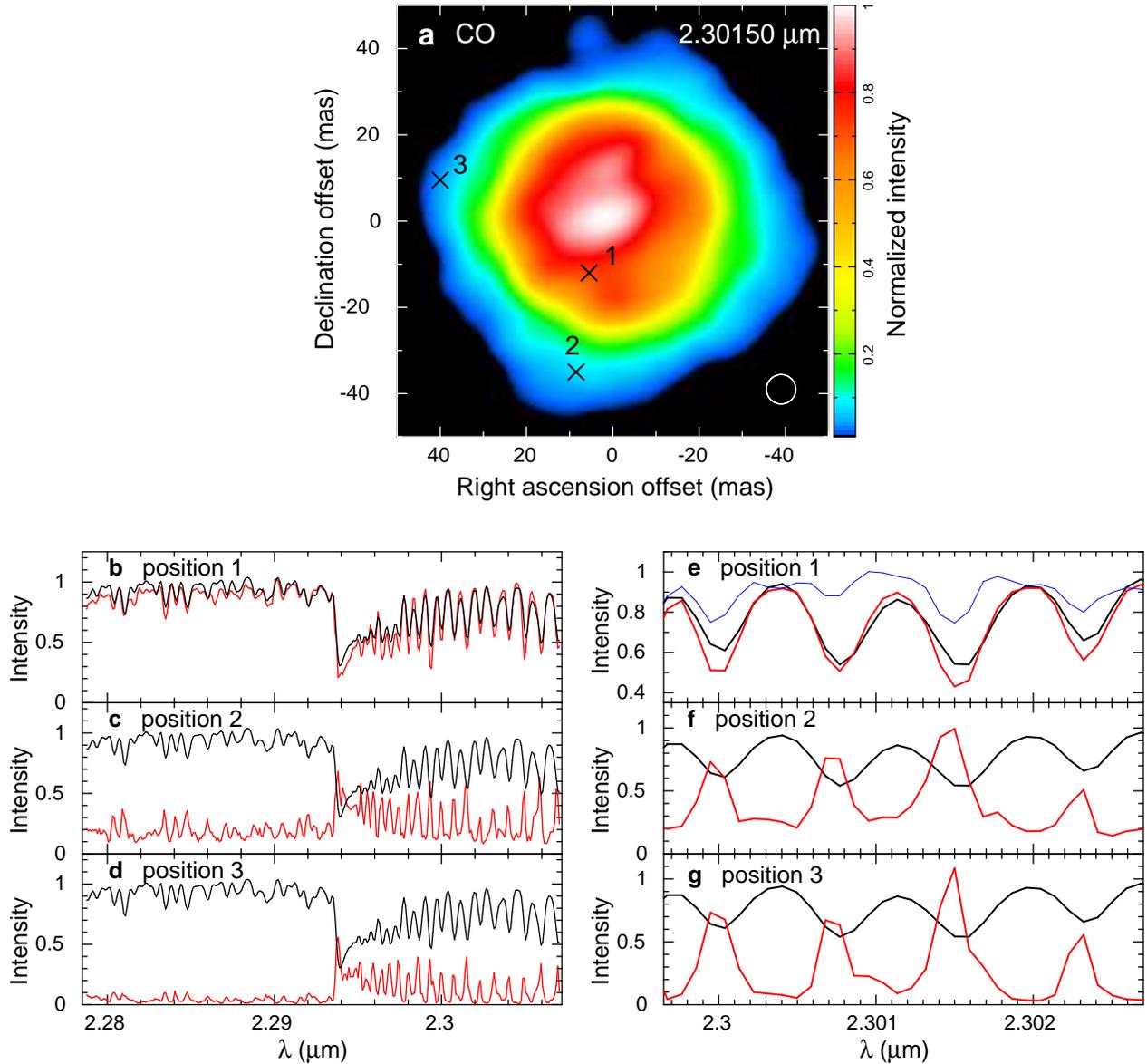}
\caption{
Spatially resolved CO line spectra obtained over the surface and 
atmosphere of R~Dor. 
{\bf a:} The image reconstructed at the center of the CO line at 
2.30150~\micron. The positions 1, 2, and 3 are where the spatially resolved 
spectra exemplarily shown in panels {\bf b}--{\bf g} were derived. 
North is up, east is to the left. 
Panels {\bf b}--{\bf d:} 
Spatially resolved spectra obtained at the positions 1, 2, 
and 3 in the entire observed spectral range are shown by the red lines. 
The spectra at the positions 2 an 3 are scaled to facilitate the visual 
inspection. 
The synthetic spectrum computed from the hydrostatic photospheric model 
is shown by the black lines. 
The spatially resolved spectra extracted off the limb of the star (panel 
{\bf c}) show not only the CO emission lines but also 
other spectral lines shortward of the CO band head, mostly due to \HOH,  
clearly in emission. 
Panels {\bf e}--{\bf g:} 
Spatially resolved spectra obtained at the positions 1, 2, 
and 3 across four CO lines are shown by the red lines.  
The synthetic spectrum computed from the hydrostatic photospheric model is 
shown by the black lines. 
In panel {\bf e}, the ratioed spectrum obtained by dividing the spatially 
resolved spectrum by the hydrostatic model spectrum is also plotted 
by the blue line to show more clearly the blueshift of the CO lines with 
respect to the model spectrum. 
The ratioed spectrum at the position 1 as well as the CO 
emission lines extracted at the positions 2 and 3 are 
blueshifted, suggesting that the gas is moving toward us. 
\label{rdor_specpos_CO}
}
\end{figure*}

Figures~\ref{rdor_specpos_CO} shows 
the spatially resolved spectra of the CO lines (red lines) at 
three positions, together with the synthetic spectrum predicted by the 
hydrostatic MARCS photospheric model (black lines). 
Figures~\ref{rdor_specpos_CO}b--\ref{rdor_specpos_CO}d show the results 
for the entire observed spectral window. 
As in the case of the \HOH\ lines described above, 
the CO lines appear in prominent emission at the 
positions 2 and 3 over the atmosphere. The other spectral lines shortward 
of the CO band head at 2.2935~\micron, mostly \HOH\ lines, also appear in 
emission at position 2 as already discussed above. 
The \HOH\ emission lines are not present in the spatially resolved spectrum 
at the position 3, because it is outside the extended \HOH\ atmosphere seen  
in Figures~\ref{rdor_images}b and \ref{rdor_images}c. 

Figures~\ref{rdor_specpos_CO}e--\ref{rdor_specpos_CO}g show an enlarged 
view of the spatially resolved spectra across four CO lines. 
For the position 1 over the surface, the ratioed spectrum (i.e., the 
spatially resolved spectrum divided by the model spectrum) is also presented 
(blue line in Figure~\ref{rdor_specpos_CO}e) 
to facilitate to see the wavelength shifts of the CO lines. 
Unlike the case of the Mg and \HOH\ lines, 
the spatially resolved spectra (as well as the ratioed spectrum) reveal 
that the CO lines are blueshifted with respect to the 
hydrostatic model spectrum. 
The images across the CO line shown in 
Figure~\ref{rdor_images} reveal that the blue-wing image 
(Figure~\ref{rdor_images}f) appears more extended than the red-wing image 
(Figure~\ref{rdor_images}i), although the flux is nearly the same 
(Figure~\ref{rdor_images}l). 
This asymmetric appearance of the image across the CO line profiles leads to 
the blueshifts in the spatially resolved spectra. 
The blueshifts of the CO lines detected over the surface and extended 
atmosphere suggest that the gas is moving toward us.

%\newpage
\subsection{Velocity-field maps over the surface and extended atmosphere}
\label{subsect_res_velmap}

The line-of-sight velocity at each position over the surface and atmosphere 
of the star was computed by taking the cross-correlation between the 
spatially resolved spectrum and the synthetic spectrum from the hydrostatic 
MARCS model (shifted to the observed wavelength frame as described above). 
To obtain a velocity-field map from the CO lines, 
the cross-correlation was computed for the lines between 2.3 and 
2.308~\micron, where the individual lines are resolved. 
For the velocity-field maps from other atomic and molecular lines, 
we selected the Mg line at 2.28164~\micron\ and 
three \HOH\ lines at 2.28043, 2.28099, and 2.28478~\micron, because they 
are mostly free of blends due to other spectral lines. 
The three \HOH\ lines have similar $gf$-values and 
lower excitation potentials, which means that they form approximately 
at the same atmospheric height. 
Therefore, we took the median of the velocity-field 
maps derived from three \HOH\ lines to obtain a more reliable result. 
Other \HOH\ lines responsible for the increases in the limb-darkened 
disk diameter and limb-darkening parameter (Figure~\ref{rdor_lddfit}) were not 
used, because their line profiles are affected by the adjacent lines. 
Other atomic and molecular lines present in the observed spectral window 
are affected by the blend and could not be used to extract velocity-field 
maps. 
We measured the line-of-sight velocity at the positions where the intensity 
is higher than 1\% of the maximum intensity of the images at the 
line center to avoid unreliable velocity measurements at very low 
intensities.

\begin{figure*}%[ht!]
\epsscale{1.16}
\plotone{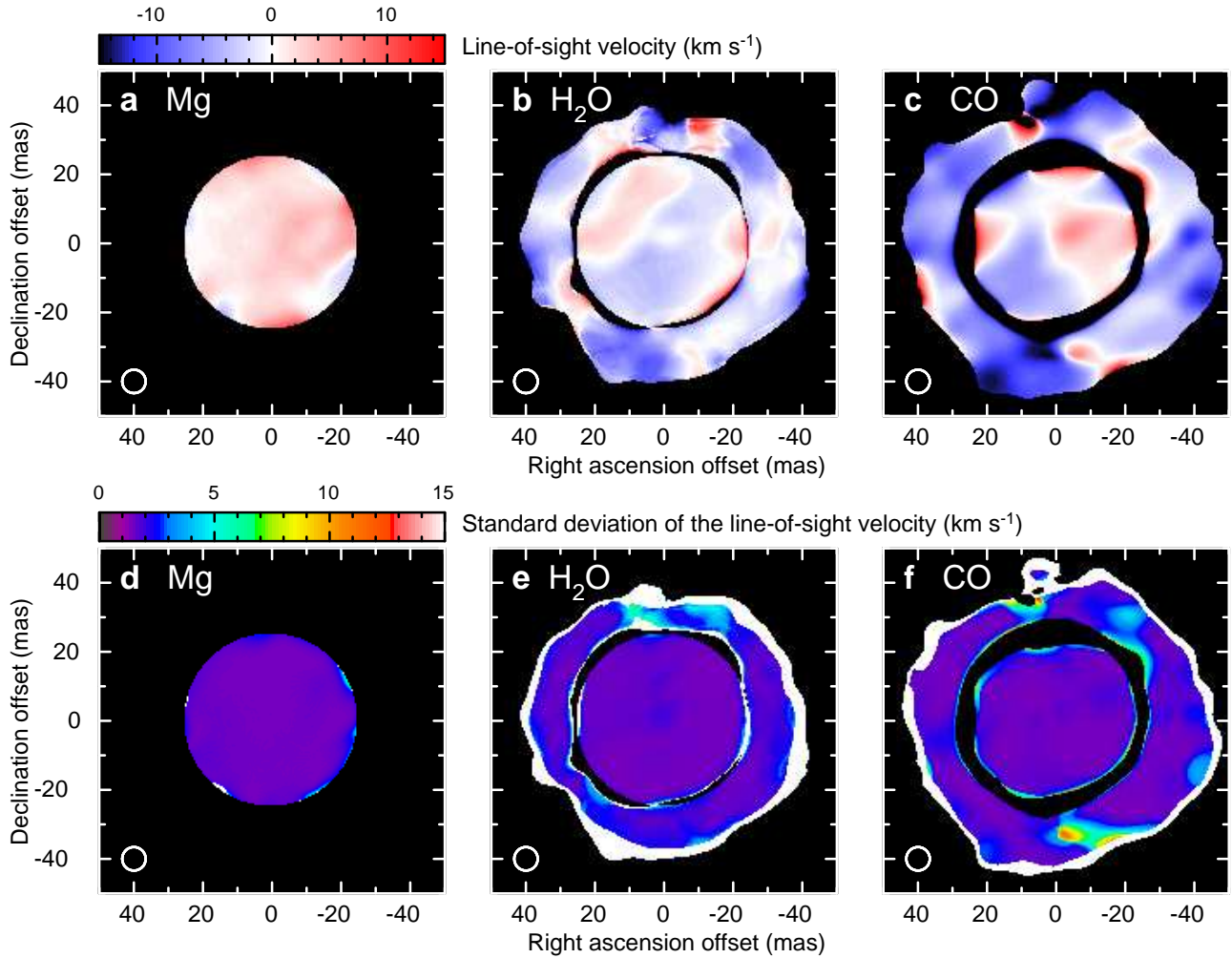}
\caption{
Velocity-field maps obtained at different atmospheric heights of R~Dor. 
North is up, east is to the left. 
{\bf a:} Velocity-field map in the deep (low) layers at 
atmospheric heights lower than $\la$1.13~\RSTAR\ (measured from the stellar 
center) obtained from the Mg line at 2.28164~\micron. 
{\bf b:} Velocity-field map at an atmospheric height of $\sim$1.5~\RSTAR\ 
obtained by averaging the velocity-field maps computed 
from three \HOH\ lines at 2.28043, 2.28099, and 2.28478~\micron. 
{\bf c:} Velocity-field map at an atmospheric height of $\sim$1.8~\RSTAR\ 
obtained from the CO first overtone lines. 
{\bf d}--{\bf f:} Standard deviation (1$\sigma$) of the line-of-sight velocity 
derived from the Mg, \HOH, and CO lines. The uncertainty in the velocity 
is at most $\sim$1.7~\KMS\ in most regions over the surface and atmosphere 
(see Appendix~\ref{appendix_reconst} for details). 
The black, ring-shaped gaps in the velocity-field maps of the \HOH\ and 
CO lines correspond to the limb of the star. 
The spatially resolved spectra on the limb show
neither absorption nor emission because the absorption and emission
cancel out within the finite beam size shown in the lower left corner.
Therefore, the velocity could not be measured. 
\label{rdor_velmap}
}
\end{figure*}

Figure~\ref{rdor_velmap} shows the maps of the line-of-sight velocity 
derived from the Mg, \HOH, and CO lines, together with the $1\sigma$ 
uncertainty in the velocity measurement (see Appendix~\ref{appendix_reconst} 
for the estimate of the uncertainty). 
As expected from the spatially resolved spectra, the line-of-sight velocity 
maps of the Mg and \HOH\ lines, shown in 
Figures~\ref{rdor_velmap}a and \ref{rdor_velmap}b, 
show that the atmospheric layers in which these lines form 
($\la$1.13~\RSTAR\ and $\sim$1.5~\RSTAR\ for the Mg and \HOH\ lines, 
respectively) are mostly quiet without significant systematic or turbulent 
motion greater than the measurement uncertainty of $\pm 1.7$~\KMS\ 
(Figures~\ref{rdor_velmap}d and \ref{rdor_velmap}e, 
see also Appendix~\ref{appendix_reconst}). 
Only in the southern and northwestern regions of the \HOH\ atmosphere do we 
detect noticeable line-of-sight velocities of about $-8$~\KMS. 
The northern region also shows large positive and negative line-of-sight 
velocities of up to $\pm 10$~\KMS. However, the uncertainty in the measurement 
is larger, $\sim$5~\KMS, as shown in Figure~\ref{rdor_velmap}e. Therefore, 
while there is some gas motion in this region, its amplitude may be less 
significant, smaller than $\pm 10$~\KMS. 

In marked contrast, the line-of-sight velocity map obtained from the CO lines, 
shown in Figure~\ref{rdor_velmap}c, 
reveals outward motion at velocities of 7--15~\KMS\ 
over a significant fraction of 
the surface and atmosphere extending out to $\sim$1.8~\RSTAR. 
The strongest outward motion at up to 15~\KMS\ is found in the southern 
region of the extended atmosphere. The southern half of the stellar disk 
adjacent to this region also shows outward motion. 
The velocity-field map derived 
from the \HOH\ lines also shows outward motion at $\sim$8~\KMS\ in this region
as mentioned above. 
The regions in the northeast and the west of the CO atmosphere 
also show outward motion at 7--10~\KMS. 
There are two small regions with positive line-of-sight velocities (i.e., moving 
away from us) in the north 
and in the south. However, as Figure~\ref{rdor_velmap}f shows, the 
uncertainty in the velocity is large at these positions, 
5--10~\KMS. Therefore, we cannot 
conclude the presence of downdrafting motion in these regions. 
Comparison of the velocity-field maps obtained from the Mg, \HOH\ 
and CO lines suggests strong acceleration of material between $\sim$1.5 
and 1.8~\RSTAR. 
This is the first three-dimensional diagnosis of the atmospheric dynamics 
from the low (i.e., deep) photospheric layers to the extended outer 
atmosphere for a star other than the Sun. 

The CO lines are estimated to be optically thick, 
if we adopt the CO column density and 
temperature estimated for M giants with stellar parameters similar to 
R~Dor \citep{ohnaka12,ohnaka19}. 
If we assume a spherically symmetric atmosphere as modeled in these previous 
works, the CO lines remain optically 
thick to the extreme limb of the CO atmosphere. 
If the outward gas motion is spherically symmetric, the line-of-sight velocity 
is expected to fall off to zero toward the edge of the extended atmosphere. 
However, it is difficult to measure the line-of-sight velocity just at 
the extreme limb of the atmosphere owing to the finite spatial resolution. 
Moreover, the line-of-sight velocity cannot be measured out to the real edge of 
the extended atmosphere, because the intensity is too low to reliably 
measure the line-of-sight velocity as mentioned above. 
In fact, as Figures~\ref{rdor_velmap}e and \ref{rdor_velmap}f show, 
the uncertainty in the velocity measurement is high at the edge of the 
\HOH\ and CO atmosphere.

Despite the blueshift over a large fraction of the surface and atmosphere, 
the spatially unresolved spectrum does not show a noticeable blueshift 
for the following reason. 
The CO lines originate not only from the extended atmosphere but also from 
the lower, static photosphere. The contribution from the outwardly moving 
extended atmosphere leads to blueshifted absorption over the stellar 
disk. However, the blueshifted emission from the extended atmosphere 
seen outside the star's limb fills in the blueshifted absorption. 
Therefore, when averaged over the surface and atmosphere, the spectrum 
does not show a noticeable blueshift.

%\newpage

The gas motion detected in R~Dor 
is very different from the dynamics of the extended 
atmosphere of the red supergiant Antares, which is characterized by 
turbulent motion of large gas clumps without any systematic outward or 
inward motion \citep{ohnaka17b}. 
The line-of-sight velocities of 7--15~\KMS\ seen in the extended atmosphere 
are higher than the terminal 
velocity of $\sim$5.5~\KMS\ that the stellar wind is supposed to 
reach at distances greater than $\sim$20~\RSTAR\ \citep{vandesande18}. 
\citet{decin18} detected velocity components much faster than 
the terminal velocity in some submillimeter molecular lines toward R~Dor, 
indicating a strong wind acceleration in addition to the slower wind. 
The outward motion seen in the line-of-sight velocity map of the CO lines may 
give rise to the high-velocity components detected in the submillimeter 
lines.

The velocity-field maps obtained from the \HOH\ and CO lines 
(Figures~\ref{rdor_velmap}b and \ref{rdor_velmap}c) show 
ring-shaped gaps along the limb of the star for the following reason. 
The spatially resolved spectra across the \HOH\ and CO lines 
along the stellar limb 
show neither absorption nor emission because the absorption and emission
cancel out within the finite beam size. Therefore, the velocity could not be 
measured along the limb of the star, which appears as the black ring-shaped 
gaps. 
We also note that 
as Figures~\ref{rdor_images}a and \ref{rdor_images}e show, the extension of 
the image in the Mg line at 2.28164~\micron\ is approximately the same as that 
of the continuum image. Therefore, the spatially resolved spectra do not 
show the Mg line in emission outside the limb of the star. This is why 
we could measure the line-of-sight velocity from the Mg line 
only inside the limb of the star as shown in Figure~\ref{rdor_velmap}a. 
\vspace*{6ex}

\section{Discussion and conclusion}
\label{sect_discuss}

The formation of clumpy dust clouds has recently been detected toward 
R~Dor, as close as at $\sim$1.5~\RSTAR, by 
polarimetric observations of light scattered by dust 
grains \citep{norris12,khouri16}. Therefore, 
dust already forms approximately at the radius of the atmosphere seen in the 
\HOH\ line images.  \citet{norris12} and \citet{khouri16} propose 
\corundum\ or Fe-free silicates such as \forsterite\ and \enstatite\ 
as grain species responsible for the 
scattered light, although their observations did not allow them to 
nail down the grain species. 

The recent three-dimensional convection simulations with dust formation 
presented by \citet{hoefner19}, although the models are more luminous and 
cooler than R~Dor, show that while \corundum\ grains form at $\sim$1.4~\RSTAR, 
\forsterite\ mantle forms onto the \corundum\ cores 
farther away, at 1.7--1.8~\RSTAR\ \citep{hoefner19}. 
Also, \citet{hoefner16} suggest that 
the radiation pressure due to scattering by \corundum\ grains is insufficient, 
and the composite grains with an \corundum\ 
core and Fe-free silicate mantle may drive the stellar wind. 
The formation radius of \corundum\ grains corresponds to 
the radius of the atmosphere probed with the \HOH\ lines 
without significant systematic motion, 
while the \corundum +Fe-free silicate grains form approximately 
at the radius of the 
atmosphere seen in the CO line images with the systematic outward motion. 
Therefore, the outward motion detected by our AMBER observations 
may be caused by the initial acceleration 
of the stellar wind driven by the radiation pressure due to scattering by 
the composite grains.

Tantalizing as the evidence is, 
it is also possible that the outward motion is simply part of 
the ballistic atmospheric motion caused by stellar pulsation or convection,
not necessarily leading to further acceleration of material. 
The three-dimensional convection simulations show the ballistic gas motion 
at $\sim$2~\RSTAR\ \citep{freytag17}. 
However, while dust formation is taken into account in these models, 
the dynamical effects triggered by the radiation pressure on dust grains 
are not yet included because it would be computationally very exhaustive. 
It is not clear either whether or not the outward motion is always present. 
The submillimeter imaging of R~Dor with a resolution 
of $30 \times 42$~mas reported by \citet{vlemmings18} using the 
Atacama Large Millimeter/submillimeter Array (ALMA) shows rotation 
at a velocity of 1.0~\KMS\ but no signature of outward motion within 100~mas 
(= 4~\RSTAR), 
which may indicate time variation in the dynamics in the outer atmosphere. 

Interferometric imaging monitoring, together with high angular 
resolution thermal-infrared imaging of the dust formation region, 
is a next step to draw 
a definitive conclusion about whether or not the material is driven outward 
by the radiation pressure on dust grains, and if so, which grain species is 
responsible for the acceleration of the stellar wind.

\acknowledgments

We thank the ESO VLTI team for their support for our VLTI/AMBER 
observations. 
K.O. acknowledges the support of the 
Comisi\'on Nacional de Investigaci\'on Cient\'ifica y Tecnol\'ogica
(CONICYT) through the FONDECYT Regular grant 1180066. 
This research made use of the \mbox{SIMBAD} database, 
operated at the CDS, Strasbourg, France, 
and NSO/Kitt Peak FTS data on the Earth's telluric features 
produced by NSF/NOAO.
We acknowledge with thanks the variable star observations from the AAVSO
International Database contributed by observers worldwide and used in this
research.


\vspace{5mm}
\facilities{VLTI(AMBER), AAVSO}


\software{
amdlib \citep{tatulli07,chelli09},
MiRA \citep{thiebaut08}
}



\clearpage
\appendix
\renewcommand{\thetable}{\Alph{section}\arabic{table}}
%\thetable
\setcounter{table}{0}

\section{Summary of AMBER observations}

The journal of our AMBER observations of R~Dor is given in
Table~\ref{obslog}.

\begin{table*}[!h]
\begin{center}
\caption{Summary of VLTI/AMBER observations of R~Dor. 
$B_{\rm p}$: Projected baseline length. 
PA: Position angle of the baseline vector projected onto the sky. 
$s$: Seeing in the visible. 
$\tau_0$: Coherence time in the visible.  
DIT: Detector Integration Time. 
$N_{\rm f}$: Number of frames in each exposure. 
$N_{\rm exp}$: Number of exposures. 
\label{obslog}
}
%\vspace*{1ex}

\begin{tabular}{r l c c l l l}\hline
\# & $t_{\rm obs}$ & $B_{\rm p}$ & PA  & $s$ & $\tau_0$ & DIT $\times N_{\rm  f} \times N_{\rm exp}$ \\
   &  (UTC)       &      (m) &($^{\circ}$)& ($^{\prime \prime}$) & (ms) & (ms) \\
\hline
\multicolumn{7}{c}{2013 Dec 01 (UTC) D0-G1-H0} \\
\hline
   1 & 02:18:39 & 63.95/57.20/55.39 &  46/172/102 &   1.2 &   1.6 & $120 \times  500 \times  5$ \\
   2 & 03:00:29 & 63.96/57.56/58.92 &  55/177/111 &   0.9 &   2.0 & $120 \times  500 \times  5$ \\
   3 & 03:43:25 & 63.63/57.56/61.95 &  64/$-$177/119 &   1.1 &   1.7 & $120 \times  500 \times  5$ \\
   4 & 04:27:05 & 62.85/57.17/64.47 &  74/$-$171/127 &   1.1 &   1.7 & $120 \times  500 \times  5$ \\
   5 & 05:18:07 & 61.42/56.28/66.59 &  84/$-$165/136 &   1.0 &   1.8 & $120 \times  200 \times  4$ \\
   6 & 05:52:58 & 59.83/55.21/67.81 &  92/$-$160/143 &   1.2 &   1.6 & $120 \times  500 \times  5$ \\
   7 & 06:33:52 & 57.74/53.70/68.78 & 102/$-$155/151 &   0.9 &   2.1 & $120 \times  500 \times  5$ \\
   8 & 07:15:54 & 55.28/51.74/69.47 & 112/$-$149/159 &   1.0 &   1.9 & $120 \times  500 \times  5$ \\
   9 & 07:56:42 & 52.71/49.42/69.84 & 122/$-$144/167 &   1.3 &   1.5 & $120 \times  500 \times  5$ \\
  10 & 08:40:21 & 50.04/46.46/70.06 & 134/$-$139/176 &   1.2 &   1.6 & $120 \times  500 \times  5$ \\
\hline
%\hline
\multicolumn{7}{c}{2013 Dec 02 (UTC) G1-H0-I1} \\
\hline
  11 & 01:40:43 & 44.45/28.01/56.63 &  16/$-$64/$-$13 &   1.6 &   1.7 & $120 \times  500 \times  5$ \\
  12 & 02:20:04 & 44.05/30.01/57.26 &  23/$-$57/ $-$8 &   1.2 &   2.3 & $120 \times  500 \times  5$ \\
  13 & 03:00:38 & 43.44/31.79/57.57 &  31/$-$51/ $-$2 &   1.4 &   1.8 & $120 \times  500 \times  5$ \\
  14 & 03:40:32 & 42.61/33.27/57.54 &  38/$-$44/  3 &   1.1 &   2.1 & $120 \times  500 \times  5$ \\
  15 & 04:19:57 & 41.53/34.47/57.20 &  45/$-$38/  8 &   1.3 &   1.9 & $120 \times  500 \times  5$ \\
  16 & 05:00:10 & 40.15/35.44/56.51 &  52/$-$31/ 14 &   1.0 &   2.3 & $120 \times  500 \times  5$ \\
  17 & 05:41:10 & 38.40/36.20/55.45 &  60/$-$24/ 19 &   1.0 &   2.5 & $120 \times  500 \times  5$ \\
  18 & 06:21:59 & 36.32/36.75/54.02 &  67/$-$18/ 24 &   1.1 &   2.2 & $120 \times  500 \times  5$ \\
  19 & 07:01:50 & 33.96/37.09/52.24 &  75/$-$11/ 30 &   1.1 &   2.3 & $120 \times  500 \times  5$ \\
  20 & 07:42:08 & 31.27/37.27/50.05 &  83/ $-$4/ 35 &   1.4 &   1.7 & $120 \times  500 \times  5$ \\
\hline
\multicolumn{7}{c}{2013 Dec 06 (UTC) A0-B2-D0} \\
\hline
  21 & 00:21:54 &  8.47/30.23/34.89 &  58/ $-$5/  7 &   1.4 &   2.7 & $120 \times  500 \times  5$ \\
  22 & 00:58:54 &  8.95/30.27/34.78 &  68/  1/ 14 &   0.9 &   4.5 & $120 \times  500 \times  5$ \\
  23 & 01:32:49 &  9.39/30.21/34.59 &  76/  6/ 21 &   1.0 &   3.7 & $120 \times  500 \times  5$ \\
  24 & 02:06:43 &  9.80/30.04/34.32 &  84/ 12/ 27 &   1.4 &   3.0 & $120 \times  500 \times  5$ \\
  25 & 02:39:52 & 10.16/29.78/33.95 &  91/ 17/ 34 &   1.1 &   3.5 & $120 \times  500 \times  5$ \\
  26 & 03:12:53 & 10.47/29.40/33.47 &  98/ 22/ 40 &   1.3 &   2.9 & $120 \times  500 \times  5$ \\
  27 & 03:47:02 & 10.74/28.89/32.82 & 106/ 28/ 46 &   1.6 &   2.2 & $120 \times  500 \times  5$ \\
\hline
\multicolumn{7}{c}{2013 Dec 06 (UTC) A1-B2-C1} \\
\hline
  28 & 05:14:33 & 11.17/ 8.98/15.16 & 124/ 41/ 88 &   2.0 &   1.9 & $120 \times  500 \times  5$ \\
  29 & 05:49:44 & 11.25/ 8.62/14.75 & 132/ 47/ 96 &   2.3 &   1.7 & $120 \times  500 \times  5$ \\
  30 & 06:23:47 & 11.29/ 8.22/14.30 & 139/ 52/104 &   2.2 &   1.6 & $120 \times  500 \times  5$ \\
  31 & 06:57:18 & 11.31/ 7.78/13.80 & 146/ 57/112 &   2.5 &   1.5 & $120 \times  500 \times  5$ \\
  32 & 08:00:50 & 11.30/ 6.79/12.80 & 160/ 67/128 &   2.5 &   1.5 & $120 \times  500 \times  5$ \\
\hline
%\label{obslog}
\end{tabular}
\end{center}
\end{table*}

\clearpage
\addtocounter{table}{-1}
\begin{table*}%[!h]
\begin{center}
\caption {
(Continued)
}
%\vspace*{1ex}

\begin{tabular}{r l c c l l l}\hline
\# & $t_{\rm obs}$ & $B_{\rm p}$ & PA  & $s$ & $\tau_0$ & DIT $\times N_{\rm  f} \times N_{\rm exp}$ \\
   &  (UTC)       &      (m) &($^{\circ}$)& ($^{\prime \prime}$) & (ms) & (ms) \\
\hline
\multicolumn{7}{c}{2013 Dec 07 (UTC) A1-B2-C1} \\
\hline
  33 & 00:01:39 &  8.26/10.05/15.79 &  53/ $-$8/ 19 &   1.5 &   2.1 & $120 \times  500 \times  5$ \\
  34 & 00:35:04 &  8.69/10.08/15.85 &  62/ $-$3/ 27 &   1.2 &   2.5 & $120 \times  500 \times  5$ \\
  35 & 01:08:01 &  9.12/10.08/15.92 &  71/  3/ 35 &   1.6 &   2.0 & $120 \times  500 \times  5$ \\
  36 & 01:46:51 &  9.61/10.04/15.97 &  80/  9/ 44 &   1.6 &   1.9 & $120 \times  500 \times  5$ \\
  37 & 02:19:58 &  9.99/ 9.96/15.99 &  88/ 14/ 51 &   1.4 &   2.3 & $120 \times  500 \times  5$ \\
  38 & 02:53:02 & 10.32/ 9.86/15.97 &  95/ 20/ 58 &   1.1 &   3.0 & $120 \times  500 \times  5$ \\
  39 & 03:27:31 & 10.62/ 9.71/15.88 & 102/ 25/ 66 &   1.0 &   3.1 & $120 \times  500 \times  5$ \\
  40 & 04:03:34 & 10.87/ 9.50/15.71 & 110/ 31/ 74 &   0.9 &   3.4 & $120 \times  500 \times  5$ \\
  41 & 04:53:03 & 11.11/ 9.14/15.34 & 120/ 38/ 84 &   0.8 &   3.9 & $240 \times  250 \times  5$ \\
  42 & 05:25:56 & 11.21/ 8.83/14.99 & 127/ 43/ 92 &   0.8 &   3.9 & $120 \times  500 \times  5$ \\
  43 & 05:58:47 & 11.27/ 8.48/14.59 & 134/ 49/ 99 &   0.9 &   3.2 & $120 \times  500 \times  5$ \\
  44 & 06:32:43 & 11.30/ 8.06/14.11 & 142/ 54/107 &   0.9 &   3.4 & $120 \times  500 \times  5$ \\
  45 & 07:06:48 & 11.31/ 7.59/13.59 & 149/ 59/115 &   1.1 &   2.7 & $120 \times  500 \times  5$ \\
  46 & 07:39:39 & 11.30/ 7.08/13.07 & 157/ 64/124 &   1.3 &   2.2 & $120 \times  500 \times  5$ \\
  47 & 08:12:35 & 11.29/ 6.52/12.57 & 164/ 69/133 &   1.4 &   2.1 & $120 \times  500 \times  5$ \\
\hline
\multicolumn{7}{c}{2013 Dec 08 (UTC) B2-C1-D0} \\
\hline
  48 & 00:16:30 & 10.07/20.17/30.23 &  -5/ -5/ -5 &   1.0 &   2.9 & $121 \times  500 \times  5$ \\
  49 & 00:50:22 & 10.08/20.20/30.27 &   0/  1/  1 &   0.8 &   3.4 & $121 \times  500 \times  5$ \\
  50 & 01:24:41 & 10.06/20.15/30.22 &   6/  6/  6 &   0.8 &   3.7 & $121 \times  500 \times  5$ \\
  51 & 01:57:41 & 10.01/20.05/30.05 &  11/ 11/ 11 &   --- &   --- & $121 \times  500 \times  5$ \\
  52 & 02:31:08 &  9.92/19.87/29.79 &  17/ 17/ 17 &   0.6 &   4.4 & $121 \times  500 \times  5$ \\
  53 & 03:07:35 &  9.78/19.59/29.37 &  23/ 23/ 23 &   0.6 &   4.3 & $121 \times  500 \times  5$ \\
  54 & 03:41:11 &  9.61/19.25/28.86 &  28/ 28/ 28 &   0.7 &   3.5 & $121 \times  500 \times  5$ \\
  55 & 04:14:52 &  9.39/18.82/28.21 &  33/ 33/ 33 &   0.9 &   2.7 & $121 \times  500 \times  5$ \\
  56 & 04:47:54 &  9.14/18.31/27.45 &  38/ 38/ 38 &   0.8 &   3.3 & $121 \times  500 \times  5$ \\
  57 & 05:22:56 &  8.82/17.66/26.48 &  44/ 44/ 44 &   0.8 &   3.3 & $121 \times  500 \times  5$ \\
  58 & 05:57:25 &  8.44/16.92/25.36 &  49/ 49/ 49 &   0.9 &   2.8 & $121 \times  500 \times  5$ \\
  59 & 06:31:00 &  8.03/16.08/24.11 &  54/ 54/ 54 &   1.0 &   2.4 & $121 \times  500 \times  5$ \\
\hline
\multicolumn{7}{c}{2013 Dec 09 (UTC) A1-B2-D0} \\
\hline
  60 & 01:03:07 &  9.16/30.26/34.70 &  72/  3/ 17 &   1.1 &   2.3 & $121 \times  500 \times  5$ \\
  61 & 01:35:54 &  9.57/30.15/34.48 &  79/  9/ 24 &   1.3 &   2.0 & $121 \times  500 \times  5$ \\
\hline
\multicolumn{7}{c}{2013 Dec 09 (UTC) A1-C1-D0} \\
\hline
  62 & 02:39:09 & 15.98/19.78/33.80 &  57/ 19/ 36 &   1.0 &   2.4 & $121 \times  500 \times  5$ \\
  63 & 03:57:19 & 15.70/19.00/32.32 &  74/ 31/ 50 &   1.1 &   2.2 & $121 \times  500 \times  5$ \\
  64 & 05:11:02 & 15.07/17.81/30.12 &  90/ 42/ 64 &   0.9 &   2.6 & $121 \times  500 \times  5$ \\
  65 & 06:23:25 & 14.13/16.17/27.14 & 107/ 54/ 78 &   1.0 &   2.5 & $121 \times  500 \times  5$ \\
  66 & 07:33:40 & 13.04/14.11/23.54 & 124/ 64/ 93 &   1.4 &   1.8 & $121 \times  500 \times  5$ \\
  67 & 08:47:08 & 12.00/11.51/19.41 & 145/ 76/112 &   0.7 &   3.5 & $121 \times  500 \times  4$ \\
\hline
\multicolumn{7}{c}{2013 Dec 10 (UTC) A1-B2-C1} \\
\hline
  69 & 01:01:46 &  9.19/10.08/15.93 &  72/  4/ 36 &   1.6 &   1.9 & $121 \times  500 \times  5$ \\
  70 & 02:16:17 & 10.08/ 9.94/15.99 &  90/ 16/ 53 &   1.1 &   2.7 & $121 \times  500 \times  5$ \\
  71 & 03:28:54 & 10.72/ 9.64/15.83 & 105/ 27/ 69 &   1.1 &   2.8 & $121 \times  500 \times  5$ \\
  72 & 04:37:36 & 11.10/ 9.16/15.37 & 120/ 38/ 83 &   0.9 &   3.1 & $121 \times  500 \times  5$ \\
  73 & 05:44:27 & 11.27/ 8.51/14.62 & 134/ 48/ 98 &   0.6 &   4.6 & $121 \times  500 \times  5$ \\
  74 & 06:52:38 & 11.31/ 7.62/13.63 & 149/ 59/115 &   0.7 &   4.0 & $121 \times  500 \times  5$ \\
  75 & 08:01:00 & 11.29/ 6.52/12.56 & 164/ 69/133 &   0.9 &   3.4 & $121 \times  500 \times  5$ \\
\hline
%\label{obslog}
\end{tabular}
\end{center}
\end{table*}

\clearpage
\section{Image reconstruction}
\label{appendix_reconst}

\begin{figure}
\epsscale{1.0}
\plotone{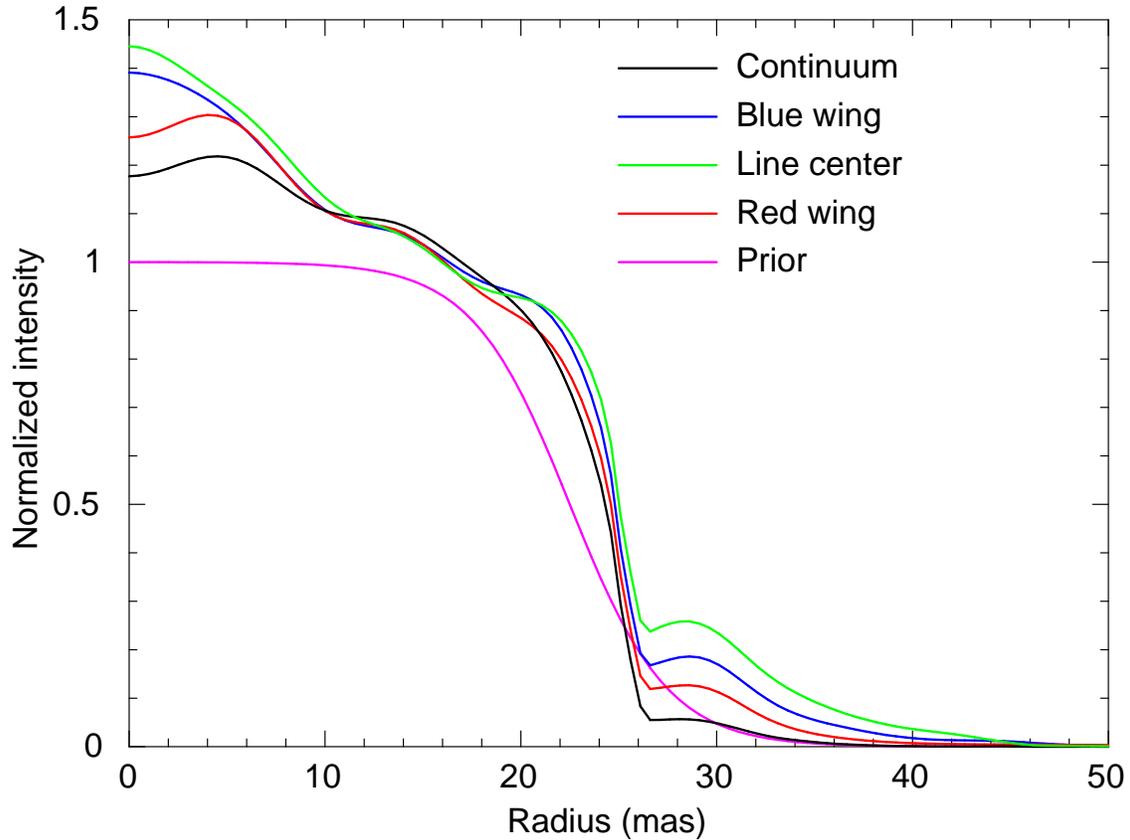}
\caption{
Radial profiles of the prior and the azimuthally averaged radial profiles 
of the images reconstructed across the CO line 
profile centered at 2.30155~\micron. 
The reconstruction was carried out with the quadratic regularization with 
$r_{\rm p} = 22.5$~mas and $\varepsilon_{\rm p} = 2.5$~mas. 
The radial profiles of the reconstructed 
images are normalized with the median of the intensity within the star's 
limb at each wavelength. 
\label{rdor_prior_profile}
}
\end{figure}

We started the image reconstruction from 
the best-fit power-law-type limb-darkened disk in the continuum with 
the limb-darkened disk diameter of 51~mas and the limb-darkening
parameter of 0.6 (Section~\ref{subsect_lddfit}). 
We took advantage of the technique that restores the 
Fourier phase from the differential phase measurements as described in 
detail in \citet{ohnaka11}. 
While satisfactory images can be reconstructed without a prior or with a 
flat prior (i.e., constant over the entire field of view of the 
reconstruction) in the continuum, 
it is necessary to apply a prior for the image reconstruction 
in the lines due to the presence of the extended atmosphere and 
more complex structures. 
As in our previous works \citep{ohnaka11,ohnaka13,ohnaka17b}, 
a Fermi-function-type prior 
$Pr(r) = 1/(\exp((r-r_{\rm p})/\varepsilon_{\rm p})+1)$ was used, where 
$r$ is the radius, $r_{\rm p}$ defines the radius where the function 
rapidly decreases to 0, and $\varepsilon_{\rm p}$ defines 
the steepness of the decrease. 
This function falls off more slowly if $\varepsilon_{\rm p}$ is 
larger, while it approaches a uniform disk (i.e., it falls off extremely 
steeply) at the limit of $\varepsilon_{\rm p} \rightarrow 0$. 
An example of 
the radial profile of the prior used in the reconstruction of the images 
presented in this paper is shown in Figure~\ref{rdor_prior_profile}. 
The Fermi-function-type prior enables us to avoid having strong artifacts 
spread over the 
entire field of the reconstruction, while allowing the presence of an 
extended atmosphere at the same time.

\begin{figure*}
%\epsscale{1.18}
\plotone{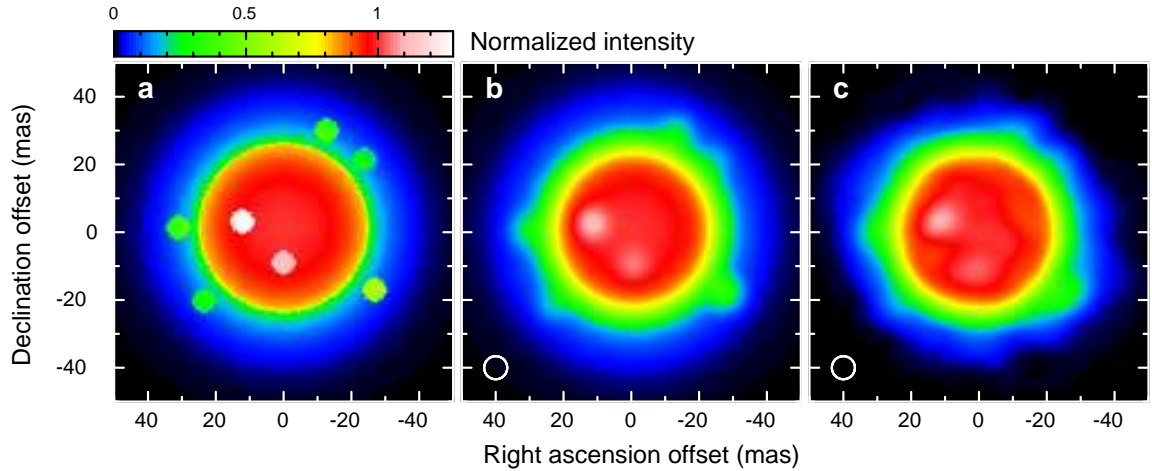}
\caption{
Image reconstruction of simulated data. 
The images are normalized with 
the intensity at the stellar disk center (i.e., the intensity higher 
than 1 means a bright spot). 
{\bf a:} Original image of the simulated star consisting of a limb-darkened 
disk, seven spots, and an extended, circular atmosphere.
{\bf b:} Image of the simulated star convolved with the same Gaussian beam 
used for the observed data.
{\bf c:} Image reconstructed from the interferometric data generated from 
the simulated stellar image for the same $uv$ coverage used in the 
AMBER observations. The reconstruction was carried out 
in the same manner as for the observed data. 
\label{simdata}
}
\end{figure*}

We examined the appropriate ranges of the prior parameters using simulated 
data. We generated an image of a star consisting of a limb-darkened 
disk and an extended, circular atmosphere with 
two spots over the surface and five spots in the atmosphere 
(Figure~\ref{simdata}a). 
We set the intensity of the extended atmosphere so that the visibility 
level of the simulated data in the first lobe matches the measurements 
in the CO lines. 
The spots over the surface have intensity contrasts of 30\% and 15\% 
with respect to the intensity at the stellar disk center. 
The spots in the atmosphere are characterized with the intensities 
of 50\%, 25\%, 17\% (two spots with this same intensity), and 12.5\% 
of the intensity at the stellar disk center. 
However, as Figure~\ref{simdata}b shows, 
after convolving with the same Gaussian beam as used for the observed 
data, the intensity contrast significantly decreases to 12\% 
(the strongest spot) to 6\% (the weakest spot). 
Interferometric observables 
(visibility amplitude, closure phase, and Fourier phase) were computed 
for the same $uv$ coverage as in our AMBER observations with noise 
comparable to the observed data. 

The image reconstruction from the simulated data 
was carried out with the quadratic (Tikhonov) 
regularization and the maximum entropy regularization using the above 
prior. 
The resulting images were convolved with the same Gaussian beam as used for 
the observed data. 
We found that the Fermi-function-type prior with $r_{\rm p}$ = 
20--25~mas and $\varepsilon_{\rm p}$ = 2.0--3.0~mas is appropriate for 
the reliable reconstruction of the original simulated image. 
Figure~\ref{simdata}c shows the image reconstructed 
with $r_{\rm p} = 22.5$~mas and $\varepsilon_{\rm p} = 2.5$~mas and the 
quadratic regularization.
The extended atmosphere 
and the spots are reasonably reproduced, although the outer edge of the 
atmosphere shows a somewhat irregular shape compared to the original data. 
If $r_{\rm p}$ or $\varepsilon_{\rm p}$ is smaller than the above 
values (i.e., the prior is more compact), the extended atmosphere of the 
simulated data cannot be well reconstructed.

We then confirmed that the Fermi-function-type prior with the above 
parameter range allows us to reconstruct images from the R~Dor data 
without producing strong artifacts spread over the field of the 
reconstruction. 
We note that the same prior was applied to all 309 spectral channels 
for a given set of $r_{\rm p}$ and $\varepsilon_{\rm p}$. 
If $r_{\rm p}$ or $\varepsilon_{\rm p}$ is larger than the above values (i.e., 
the prior is more extended or falls off more slowly), 
significant artifacts appear everywhere in the field of the 
reconstruction. This is also the case if the reconstruction {\em in the lines} 
is carried out 
without a prior or a flat prior (i.e., prior extended over the entire field 
of the reconstruction). 
The reason is that the absence of data at baselines even 
shorter than the shortest baseline of 6.5~m of the present data 
makes it difficult to reconstruct an extended component. 

We used the quadratic (Tikhonov) regularization and the maximum entropy
regularization for the image reconstruction from the R~Dor data. 
The image reconstruction was carried out with nine different combinations 
of the prior parameters with $r_{\rm p}$ = 20.0, 22.5, and 25.0 and 
$\varepsilon_{\rm p}$ = 2.0, 2.5, and 3.0 using each regularization scheme, 
resulting in 18 different set-ups in total. 
We computed the median of the images reconstructed with 18 different 
set-ups at each wavelength. 
The spatially resolved spectra over the surface and atmosphere presented 
in Section~\ref{subsect_res_spatspec} were extracted from the data cube 
of these median images. 

The image reconstruction with MiRA is carried out by comparing the 
interferometric observables from the observations and unconvolved images. 
Therefore, MiRA's output images often have a super-resolution of a factor of
3--5. However, this super-resolution of a factor of 3--5 corresponds to
baselines 3--5 times longer than the maximum baseline of the
observations. Because we do not have any directly measured information at
baselines longer than the actual maximum baseline, the super-resolution images
are partially based on extrapolation of Fourier data to longer baselines (to
reconstruct structures smaller than the diffraction limit of 
$\lambda/B_{\rm max}$). Therefore, reconstructed structures smaller than the 
diffraction limit of $\lambda/B_{\rm max}$ may be real but may also be 
artifacts. 
By convolving the MiRA's super-resolution output image with the Gaussian
beam with the FWHM of $\lambda/B_{\rm max}$, 
the artifacts are suppressed. While we may lose structures that are smaller
than the diffraction limit and may be real, we can focus structures resolved 
with the conservative diffraction-limited resolution.

\begin{figure*}
\epsscale{1.16}
\plotone{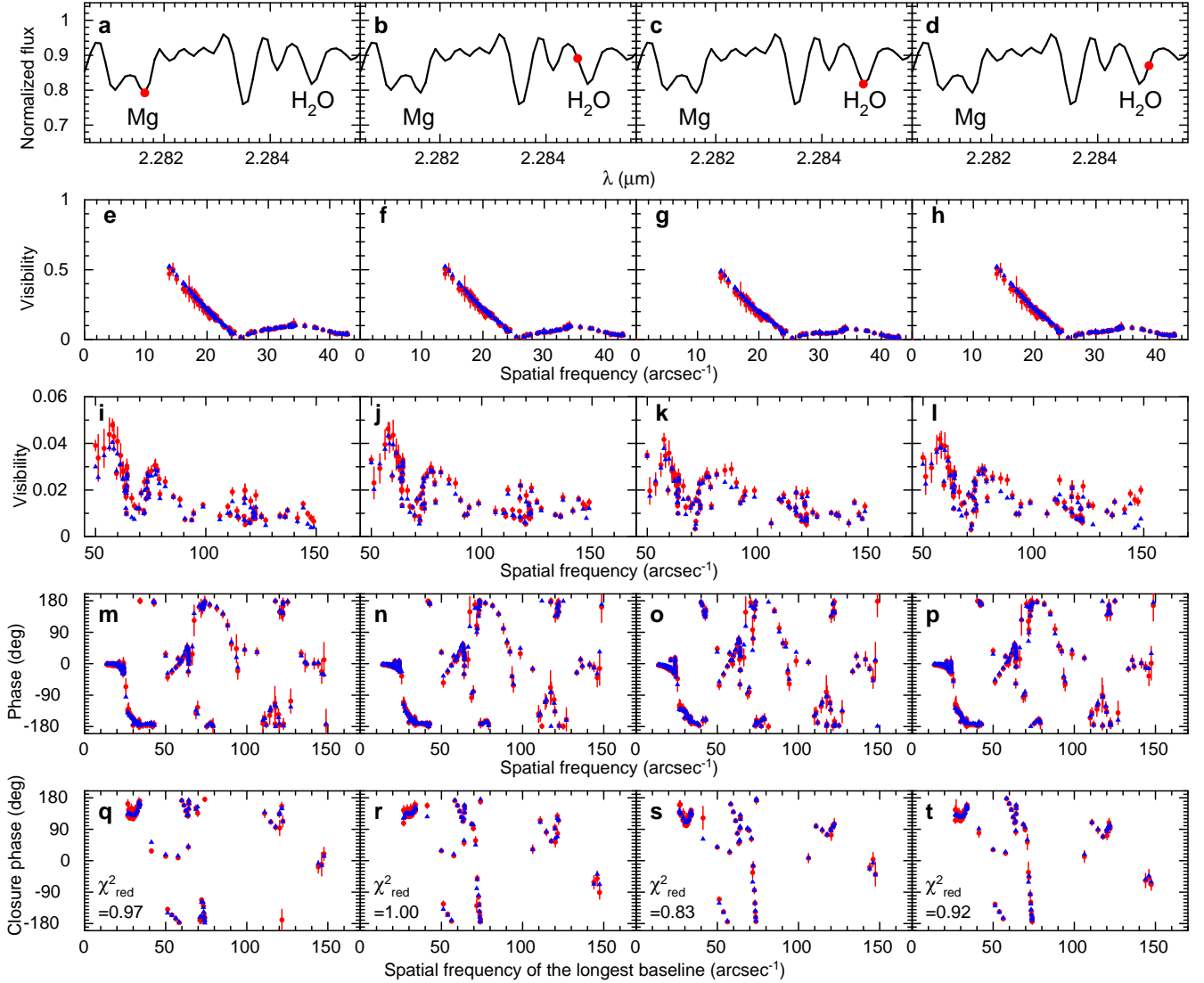}
\caption{
Comparison of the observed interferometric observables and those from the 
images reconstructed in the Mg (2.28164~\micron) and \HOH\ lines 
(2.28478~\micron). 
The reconstruction was carried out with the quadratic regularization with 
$r_{\rm p} = 22.5$~mas and $\varepsilon_{\rm p} = 2.5$~mas. 
The top row ({\bf a}--{\bf d}) shows the observed spectrum across the Mg 
and \HOH\ lines. The interferometric observables at the wavelength channels 
marked by the filled circles are shown in the corresponding columns. 
The second, third, fourth
and fifth rows show comparisons of the visibility at spatial frequencies
lower than 45~arcsec$^{−1}$, visibility at spatial frequencies higher than
45~arcsec$^{-1}$, Fourier phase, and closure phase, respectively. 
The red dots with the error bars represent the observed data, 
while the blue triangles represent the values from the reconstructed images. 
The reduced  $\chi^2$ values including the visibilities, Fourier phases and 
closure phases, are given in the bottom row.
\label{fit_weak}
}
\end{figure*}

\begin{figure*}
\epsscale{1.16}
\plotone{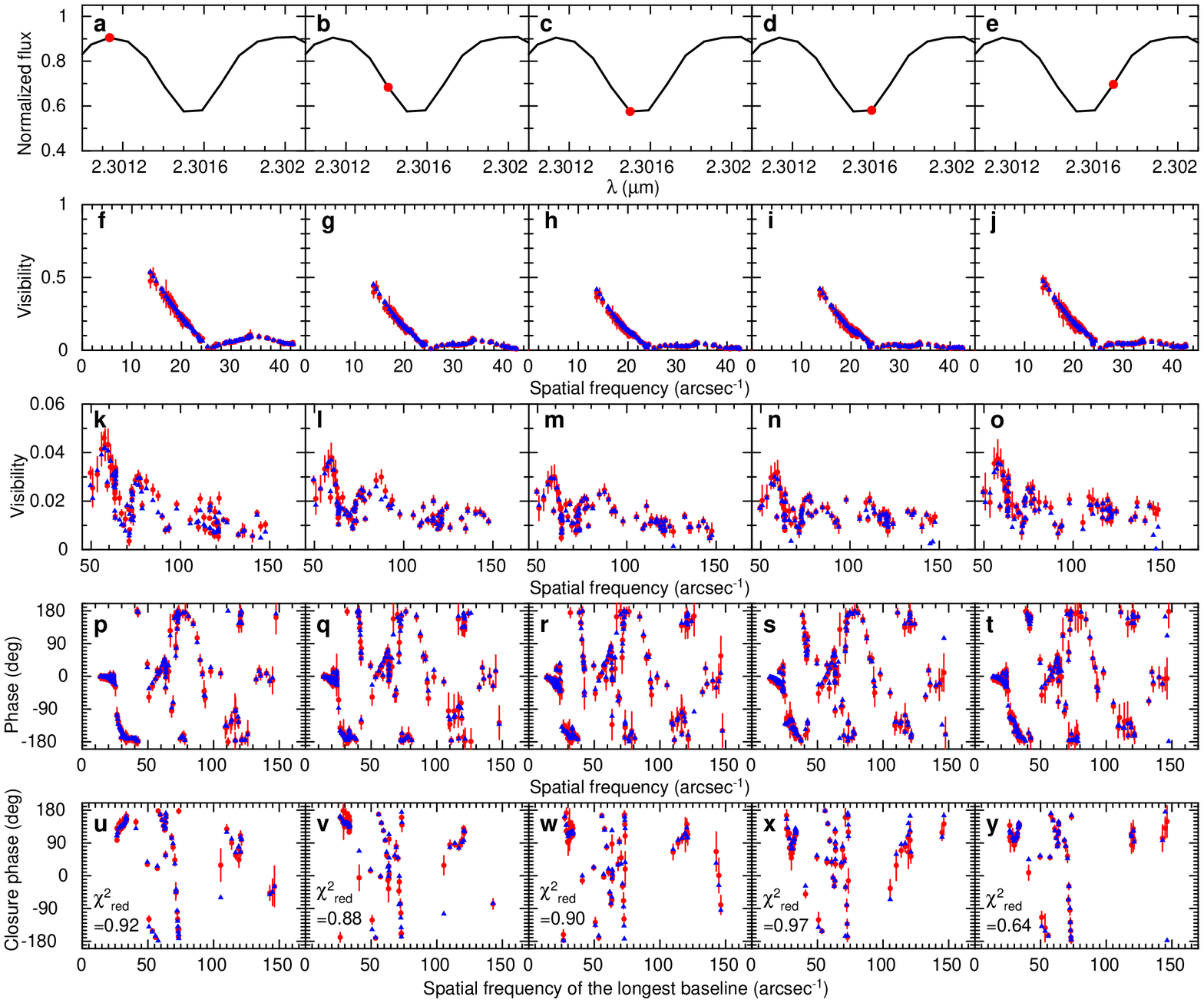}
\caption{
Same as Figure~\ref{fit_weak} but for the CO line at 
2.30150~\micron, shown in Figure~\ref{rdor_images}l. 
\label{fit_CO}
}
\end{figure*}

Figure~\ref{rdor_prior_profile} shows the radial profile of the 
Fermi-function-type prior with $r_{\rm p} = 22.5$~mas and 
$\varepsilon_{\rm p} = 2.5$~mas and the azimuthally averaged radial profiles 
of the images 
reconstructed across the CO line profile centered at 2.30155~\micron\ 
(Figure~\ref{rdor_images}l). 
The radial profiles were obtained from the 
reconstructed images not convolved with the Gaussian beam for comparison 
with the prior. 
Because the reconstructed images show significant wavelength-dependent 
surface inhomogeneities, 
we opted to normalize the radial profiles with the median of the intensity 
over the stellar disk, instead of normalizing with the maximum 
intensity. The figure shows that the spatial extension of the atmosphere 
as well as the surface structures significantly changes across the line 
profile, although the same, wavelength-independent prior was used. 
This suggests that the wavelength-dependence of the reconstructed images 
is not dominated by the prior but it reflects the observed data. 
Comparisons between the measured interferometric
observables and those from the images reconstructed at nine wavelength 
channels in the lines of Mg, \HOH, and CO (quadratic regularization with 
$r_{\rm p} = 22.5$~mas and $\varepsilon_{\rm p} = 2.5$~mas) are shown in 
Figures~\ref{fit_weak} and \ref{fit_CO}.

To estimate the uncertainty in the line-of-sight velocity measurement, 
we extracted the velocity-field maps of the Mg, \HOH, and CO lines 
from each of the data cubes reconstructed with 18 different set-ups described 
above. Then we obtained maps of the standard deviation of the line-of-sight 
velocity from the 18 velocity-field maps. 
The standard deviation of the line-of-sight velocity among the 18 maps 
is at most 1~\KMS\ in most parts over the surface and atmosphere. 
The uncertainty in the systemic velocity 
is estimated to be 1~\KMS, given the range of 6--8~\KMS\ 
(in the local standard of rest) obtained from 
various radio/far-IR observations as discussed in Sect.~\ref{subsect_marcs}. 
Adding the uncertainty in the wavelength calibration (1.0~\KMS), 
the total $1\sigma$ uncertainty in the velocity measurement is estimated to 
be 1.7~\KMS. 
Figures~\ref{rdor_velmap}d--\ref{rdor_velmap}f show the maps of the total 
standard 
deviation (1$\sigma$) including the uncertainty originating from 18 different 
reconstruction set-ups and the uncertainty in the systemic velocity and 
the wavelength calibration. 

The data files of the interferometric observables used for our image 
reconstruction are available in the OIFITS format at the Optical
interferometry DataBase (OiDB) at the Jean-Marie Mariotti Center 
(http://oidb.jmmc.fr/index.html).

\end{document}